\documentclass{jfm}
\usepackage{graphicx}
\usepackage{subfig}

\usepackage{amsmath}

\def\Rey {{Re}}

\def\reffig#1{Fig.~\ref{fig:#1}}

\def\vel {\mathbf{u}}

\shorttitle{Self-sustained LSMs in ASBL}
\shortauthor{S. Azimi, C. Cossu and T. M. Schneider}

\title{Self-sustained large-scale motions in the asymptotic suction boundary layer}

\author{Sajjad Azimi\aff{1},
Carlo Cossu\aff{2}
\and Tobias M.  Schneider\aff{1}
\corresp{\email{tobias.schneider@epfl.ch}}}

\affiliation{\aff{1}Emergent Complexity in Physical Systems Laboratory (ECPS), \'Ecole Polythechnique F\'ed\'erale de Lausanne, CH-1015 Lausanne, Switzerland
\aff{2}
LHEEA, UMR 6598  CNRS Centrale Nantes, F-44300 Nantes, France}

\begin{document}

\maketitle

\begin{abstract}
Large-scale motions, also known as superstructures, are dynamically relevant coherent structures in a wall-bounded turbulent flow, that span the entire domain in wall-normal direction and significantly contribute to the global energy and momentum transport. 
Recent investigations in channel and Couette flow, suggest that these large-scale motions are self-sustained, implying they are not driven by small-scale motions at the wall. 
Whether large-scale motions are self-sustained has however not yet been answered for open boundary layers, which are relevant for many applications. 
Here, using the asymptotic suction boundary layer flow at the friction Reynolds number $Re_\tau = 1\,168$ as a testbed, we show that large-scale motions are self-sustained also in boundary layers. 
Together with the previous investigations in confined flows, this observation provides strong evidence of the robust and general nature of coherent self-sustained processes in turbulent wall-bounded flows.
The dynamics of the large-scale self-sustained process involving the growth, breakdown and regeneration of quasi-streamwise coherent streaks and vortices within the boundary layer shows temporal phase relations reminiscent of bursting as observed in buffer layer structures. The dynamics however differs from the quasi-periodic large-scale streak-vortex regeneration cycle observed in confined flows. Based on the similarity of the dynamics of large-scale motions in boundary layers and of small-scale buffer layer structures, we conjecture that the bursting behaviour is associated with the dynamical relevance of only one wall, while two confining walls lead to the quasi-periodic cycle. 
\end{abstract}


\section{Introduction}
Wall bounded turbulence is characterised by coherent structures in a wide range of scales \citep{Townsend1980}. These structures range from small-scale streaky motions in the near-wall region \citep{Kline1967} to large-scale motions with the size of the geometrical constraint of the turbulent flow \citep{Kovasznay1970,Komminaho1996,Kim1999,Hutchins2007}.
Large-scale and very-large scale motions, also known as superstructures, carry a significant fraction of the turbulent kinetic energy and contribute significantly to the Reynolds shear stress \citep{Guala2006}.

Despite their importance for global momentum transport, there is no general consensus on the mechanism underlying the generation and the sustaining of large-scale motions in wall-bounded turbulent flows.
A widespread interpretation relates the formation of the large-scale motions to scale-growths mechanisms of hairpin vortices which are themselves fed by the active buffer layer streaky structures \citep{Kim1999,Tomkins2003}. Based on these ideas, large-scale motions could not exist independently of the near-wall active small-scale streaky structures.
However, an increasing number of numerical and experimental studies suggest that large-scale motions might be generated and sustained independently of driving due to the near-wall active small-scale motions:
\cite{Flores2006a} and \cite{Flores2007a} show that the dynamics of large-scale motions is not significantly affected when the buffer-layer structures are destructed by means of wall roughness.
This observation implies that buffer-layer active structures are not necessarily the only structures feeding large-scale motions.
\cite{Pujals2009,Hwang2010a,Hwang2010b} and \cite{Willis2010} show that large-scale motions efficiently extract energy directly from the turbulent mean flow via non-modal energy amplification mechanisms. Large-scale motions could therefore be self-sustained without feeding by smaller-scale hairpin structures.
\cite{Hwang2010,Hwang2011} and \cite{Rawat2015a} demonstrate that indeed, both large-scale and log-layer coherent motions can be self-sustained. They show that these motions survive in both channel and Couette flow, when smaller-scale active motions are artificially quenched and replaced by purely dissipative structures by means of overfiltered large-eddy simulations (LES). Using the same overfiltered LES approach \cite{Rawat2015a} and \cite{Hwang2016a} have been able to compute invariant coherent large-scale solutions of the (LES) filtered Navier-Stokes equations.
Building on these results, \cite{Cossu2017} suggest that Townsend's attached eddies \citep{Townsend1976}, believed to be the skeleton of wall-bounded turbulence  \citep{Marusic2019}, consist of quasi-streamwise \emph{coherent} streaks and vortices which are self-sustained by a mechanism similar to the one sustaining transitional \citep{Boberg1988,Waleffe1995} and buffer-layer coherent structures \citep{Jimenez1991,Hamilton1995}.

To date, evidence of the self-sustained nature of coherent large-scale motions has been provided only for the (internal) parallel pressure driven channel \citep{Hwang2010,Hwang2011} and Couette \citep{Rawat2015a} flows at relatively low Reynolds numbers ($\Rey_\tau \simeq 550$ and $\Rey_\tau \simeq 128$ respectively). Related attempts in Hagen-Poiseuille flow have been inconclusive \citep{Feldmann2018}.
Further evidence that large-scale motions are generically self-sustained is therefore needed, especially for high Reynolds number boundary layer flows such as those relevant for atmospheric dynamics or wind engineering and vehicle external aerodynamics, where large-scale motions greatly affect performance.
Recent results of \cite{Kevin2019a,Kevin2019} indicate that the features of large-scale motions in experimentally studied turbulent boundary layers are consistent with those of a coherent self-sustained process. There is however no direct evidence that large-scale motions are self-sustained also in boundary layers. 

In this study we investigate if large-scale motions are self-sustained in boundary layers at high Reynolds numbers. We follow the overfiltered large-eddy simulation (LES) approach \citep{Hwang2010,Hwang2011,Rawat2015a,Hwang2016a} where active small-scale structures are removed from the flow by increasing the width of the spatial filter in an LES while keeping the numerical grid constant. To avoid the difficulties related to both the non-parallel nature of growing turbulent boundary layers and to their strong local sensitivity to the upstream flow, 
we consider the asymptotic suction boundary layer (ASBL) flow \citep{Schlichting2004}. In ASBL, constant suction through the wall arrests the growth of the boundary layer thickness yielding parallel flow conditions. 
The kinetic energy associated with large-scale motions of turbulent flow in ASBL is relatively weak when compared to other flow systems \citep{Schlatter2011, Bobke2016}. This is evidenced by the ratio of the large-scale to the small-scale peaks in the energy spectrum being small compared to channel or Couette flow, where stronger large-scale motions are observed.  
Due to the relative strength of the small-scale structures, ASBL is a particularly appropriate flow to examine the self-sustained nature of large-scale motions. If the large-scale motions are found to be self-sustained even in the presence of the especially energetic near-wall small-scale structures of ASBL, such a finding would point towards a robust and generic self-sustaining mechanism of large-scale coherent motions in boundary layers.

The structure of the paper is as follows. In section \ref{sec:methods}, we introduce the flow system, discuss the governing equations and specify numerical methods for solving those equations. In section \ref{sec:results} we first introduce a modified overfiltering approach that allows to successfully isolate large-scale motions from the dynamics of the damped  near-wall small-scale structures. We thereby show that that large-scale motions in ASBL are self-sustained. Based on an analysis of one isolated large-scale motion, we describe the self-sustained mechanism and discuss its properties in comparison to similar processes reported in other flow systems. The results are discussed in the concluding section \ref{sec:conclusion}.

\section{Methodology}\label{sec:methods}

\subsection{Asymptotic suction boundary layer flow (ASBL)}

\begin{figure}
    \centering
    \includegraphics[width=0.5\textwidth]{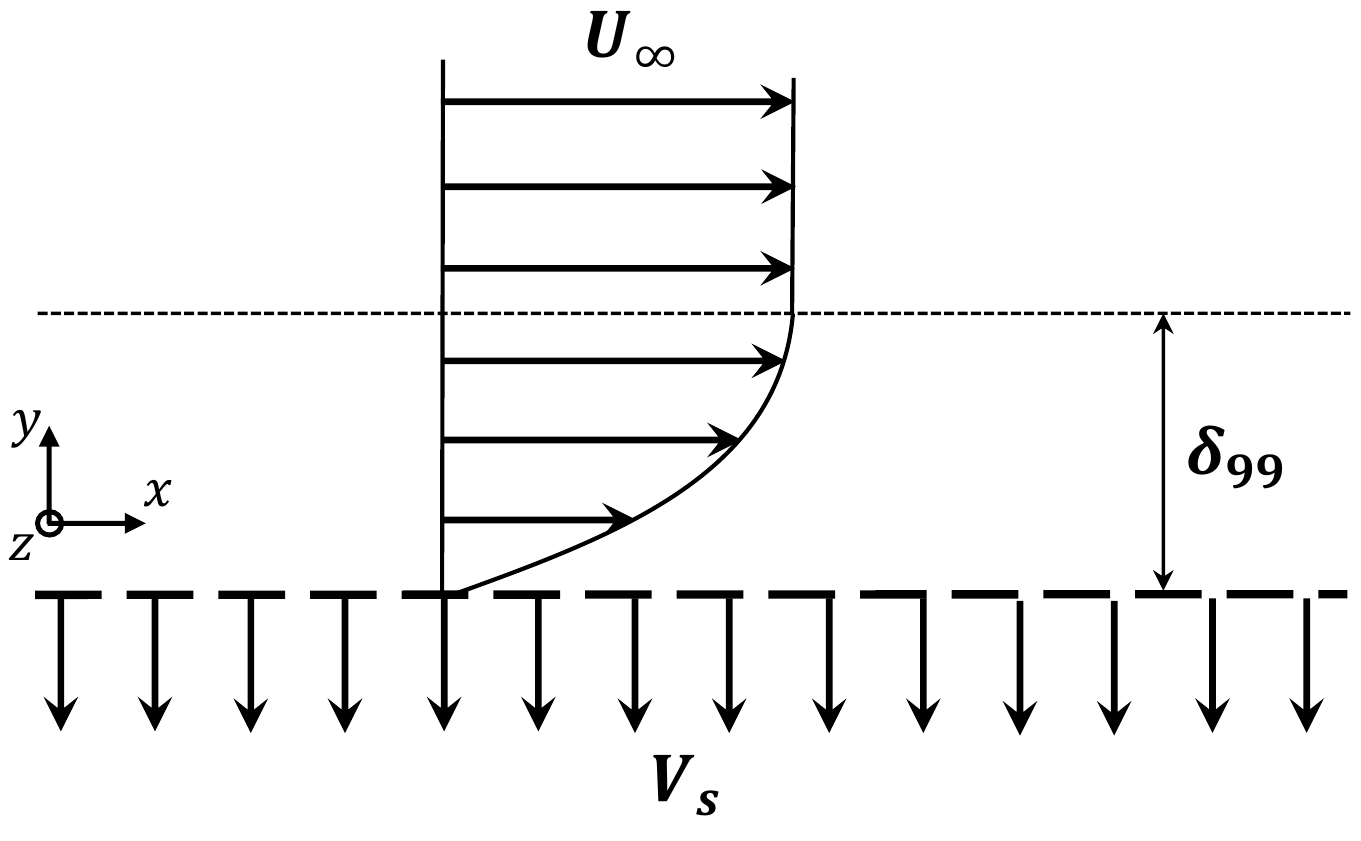}
    \caption{Schematic of asymptotic suction boundary layer flow. The turbulent boundary layer thickness is $\delta_{99}$, the height where the mean streamwise velocity reaches $99\%$ of the free stream velocity $U_\infty$. The value of the Reynolds number $\Rey=U_\infty/V_s$, given by the ratio of the free steam velocity and the uniform suction velocity $V_s$, is fixed to $\Rey=300$.
    }
    \label{fig:schematic}
\end{figure}

We consider the flow of a uniform free stream of velocity $U_\infty$ over a flat plate with uniform and constant wall-normal suction $V_s$. Far from the plate's leading edge, the growth of the boundary layer thickness is compensated by the wall suction and asymptotic suction boundary layer flow (ASBL) is reached (see \reffig{schematic}). Here the downstream momentum that enters the boundary layer from the free stream exactly balances wall friction so that the boundary layer remains parallel.  
ASBL allows for a laminar solution of the form $\Check{U}_l=U_\infty \left(1-\exp \left(-\Check{y}/\delta\right)\right)$, where $\check{U}_l$ is the (dimensional) streamwise velocity, $\check{y}$ the (dimensional) wall-normal coordinate and $\nu$ is the kinematic viscosity of the fluid.
We choose to non-dimensionalise the problem with the displacement thickness of the laminar solution $\delta^* = \nu / V_s$ as a length scale and the free-stream velocity $U_\infty$ as velocity scale. Time is measured in units of $\delta^*/U_\infty$. The flow has a single control parameter, namely the Reynolds number $\Rey=U_\infty (\nu/V_s)/\nu=U_\infty/V_s$ based on the free-stream velocity and the laminar displacement thickness. 

Properties of wall-bounded turbulence in the near-wall region are universal when measured in inner units of the flow. Momentum balance in ASBL allows us to express the inner velocity scale $u_\tau$, and the inner length scale $\delta_\tau$ relative to the outer units used for non-dimensionalisation in terms of the externally controlled Reynolds number, $u_\tau=U_\infty/\sqrt{Re}$ and $\delta_\tau=\delta^*/\sqrt{Re}$, respectively.
The friction Reynolds number $Re_\tau$, that measures the scale separation between the characteristic length scale of the large-scale motions $\delta_{99}$ and the characteristic length scale of the near-wall small-scale structures $\delta_\tau$, is equal to $Re_\tau = \delta_{99} \sqrt{Re} / \delta^*$.
Throughout this paper, all variables in inner units are denoted by a superscript plus sign.

ASBL is linearly stable up to $\Rey \approx 54370$ \citep{Hocking1975}.
In practice, for $\Rey > 270$ the flow is turbulent \citep{Khapko2016}.
ASBL is characterised by high scale separations even at the smallest Reynolds numbers where turbulence is sustained \citep{Khapko2016}.
The kinetic energy associated with the large-scale motions is relatively weak compared to the energy of the small-scale streaky structures in the near-wall region \citep{Schlatter2011, Bobke2016}.

\subsection{Governing equations}
We consider the flow evolution under the filtered Navier-Stokes equations \citep[see e.g.]{Pope2000}.
These equations underlying LES describe the evolution of the filtered velocity, namely, the velocity contributions of spatial scales larger than a chosen filter width.
With the streamwise, wall-normal and spanwise coordinates denoted by $\mathbf{x}=[x,y,z]$, respectively and the corresponding non-dimensional total velocity components indicated by $\vel=[u,v,w]$, the governing equations for the filtered velocity are
\begin{equation}
\begin{aligned}\label{eq:filtered_eq}
    \frac{\partial \overline{u}_i}{\partial t} + \overline{u}_j\frac{\partial \overline{u}_i}{\partial x_j} &= -\frac{\partial \overline{q}}{\partial x_i} + \nu \frac{\partial^2 \overline{u}_i}{\partial x_j^2} - \frac{\partial \overline{\tau}_{ij}^r}{\partial x_j}\\
    \frac{\partial \overline{u}_i}{\partial x_i} &= 0.
\end{aligned}
\end{equation}
The action of the filter is denoted by an overbar. The influence of the scales smaller than the filter width on the filtered velocity is captured by the residual stress tensor
 $\overline{\tau}^r = \overline{\tau}^R-tr(\overline{\tau}^R)\mathbf{I}/3$, with $\overline{\tau}^R=\overline{u_i u_j}-\overline{u_i}\,\overline{u_j}$ and $\overline{q} = \overline{p}+tr(\overline{\tau}^R)/3$. 
The residual stress depends on the total unfiltered velocity field. To close the equations, the residual stress tensor $\overline{\tau}^r$ thus needs to be modelled and expressed in terms of the filtered velocity. We choose the static \cite{Smagorinsky1963} model where the residual stress tensor is given by $\overline{\tau}^r_{ij}=-2\nu_t S_{ij}$ where $S_{ij}$ is the rate of the strain tensor of the filtered velocity field and $\nu_t$ is the eddy viscosity. $\nu_t$ is given by
\begin{align*}
    \nu_t=D(C_s\overline{\Delta})^2 \overline{S},
\end{align*}
with $\overline{S}=\sqrt{2\overline{S}_{ij}\overline{S}_{ij}}$ and $\overline{\Delta}=\sqrt[3]{\Delta x \Delta y \Delta z}$ given in terms of the grid spacing in all three directions. 
The wall damping function $D=1-\exp \left(-\left( y^+ /A^+ \right)^3\right)$ ensures the residual stress to be zero at the wall. Following \cite{Kim1999a} we choose $A^+=25$. 
The only parameter varied, is the Smagorinsky constant $C_s$ which controls the filter width \citep{Mason1986} and thereby the strength of the filtering. 
We often consider values of $C_s$ larger than reference values typically used in an LES that attempts to reproduce DNS results. In a standard LES, the filter width is adapted to the resolution of the numerical grid. Here, we instead follow the overfiltering approach of \cite{Hwang2010}, where the numerical grid resolution is chosen fine enough to resolve buffer-layer structures. Increasing $C_s$ beyond its reference value allows us to explicitly filter out an increasingly large range of scales that could be resolved using the numerical grid. 

In contrast to other more elaborate subgrid models used in LES simulations, the \emph{static} \cite{Smagorinsky1963} model prevents backscatter of energy from small-scale structures to large-scale motions but captures dissipation at small scales. This is key to investigate the self-sustained nature of large-scale motions by isolating them. If large-scale motions are sustained while small-scale active structures are quenched by the overfiltering, and there is no energy flux from the quenched scale to larger scales, the large-scale motions are self-sustained.  

Energy is injected into the boundary layer at the constant rate of $I = V_s U^2_\infty /2$ per area of the plate. For statistically stationary turbulence, the time-averaged energy dissipation $D$ equals the energy input. Consequently, the rate at which energy is dissipated in ASBL  depends only on the free stream and suction velocity but is independent of $C_s$. Therefore, large-scale motions isolated by overfiltered simulations represent the physically correct energy input and dissipation rate at any given Reynolds number. 

\subsection{Numerical setup}

\begin{table}
    \centering
    \begin{tabular}{cccccccc}
    \hline
    \hline
    Name & $L_x$ & $H$ & $L_z$ & $N_x$ & $N_y$ & $N_z$ & $Section$ \\
    \hline
    \hline
    RefDNS & $256$ & $150$ & $128$ & $256$ & $241$ & $256$ & \ref{subsec:refCase}\\
    RefLES & $256$ & $100$ & $128$ & $64$ & $61$ & $64$ & \ref{subsec:refCase}\\
    LESbox & $512$ & $100$ & $256$ & $128$ & $61$ & $128$ & \ref{subsec:ODL_SSM} -- \ref{subsec:ODL_EMM}\\
    LSMbox & $192$ & $100$ & $84$ & $48$ & $61$ & $42$ & \ref{subsec:dynamics} \& appendix \ref{sec:ODL_EMM_LSMbox}\\
    \hline
    \hline
    \end{tabular}
    \caption{Parameters of the numerical domains used in this study. Given is the size of the domain $[L_x,H,Lz]$ and the number of grid points before dealiasing $[N_x,N_y,N_z]$. RefDNS and RefLES are used to determine a reference value of $C_s$ for which the LES reproduces DNS results. Overfiltered simulations are carried out in the LESbox, while the detailed dynamics of a single isolated large-scale motion is studied in the LSMbox.}
    \label{tab:domains}
\end{table}

To study the self-sustained nature of large-scale motions in ASBL we apply the overfiltering approach at $Re=300$ corresponding to a friction Reynolds number $Re_\tau = 1168$.
For our simulations, we consider a numerical domain of length $L_x$, width $L_z$ and height $H$, where periodic boundary conditions are applied in streamwise $x$ and spanwise $z$ directions. Dirichlet boundary conditions are enforced on the wall $y=0$ as well as on a top plane at $y=H$ parallel to and sufficiently far from the wall to approximate the semi-infinite space,
    $\overline{\vel}(x,0,z) = [0,-1/Re,0];\, \overline{\vel}(x,H,z) = [1,-1/Re,0]$.
The box height $H$ must be chosen sufficiently large so that flow structures completely detach from the top plane and results become independent of $H$. 
To close the problem, we impose zero mean pressure gradient in streamwise and spanwise directions.

The governing equations~(\ref{eq:filtered_eq}) are integrated in time using an extension of the code Channelflow 2.0 \citep{Gibson2019}.
Channelflow implements a pseudo-spectral method using a spectral Fourier-Chebychev-Fourier discretisation in streamwise, wall-normal and spanwise directions, respectively. A third order accurate semi-implicit backward differentiation method is employed to advance the equations in time.
The 2/3 dealiasing rule is applied in streamwise and spanwise directions.

The domain is discretised with $[N_x,N_y,N_z]$ collocation points in streamwise, wall-normal and spanwise directions, respectively. Specific resolutions of the different domains used are provided in table \ref{tab:domains}.
The chosen discretisations correspond to two different resolutions of the numerical grid: A fine resolution is used to accurately resolve all the scales of the flow in a direct numerical simulation, DNS.
For large-eddy simulations, LES, we consider a second coarser resolution to reduce the computational cost of the simulations. Parameters of the chosen grid resolutions are summarised in table~\ref{tab:resolution}.

\begin{table}
    \centering
    \begin{tabular}{ccccccccc}
    \hline
    \hline
    Grid & $\Delta x$ & $\Delta y_{min}$ & $\Delta y_{max}$ & $\Delta z$ & $\Delta x^+$ & $\Delta y_{min}^+$ & $\Delta y_{max}^+$ & $\Delta z^+$ \\
    \hline
    \hline
    DNS-grid & $1.5$ & $0.04$ & $0.98$ & $0.5$ & $26$ & $0.67$ & $21.2$ & $13$\\
    LES-grid & $6.0$ & $0.13$ & $2.62$ & $3.0$ & $104$ & $2.2$ & $45.3$ & $52$\\
    \hline
    \hline
    \end{tabular}
    \caption{Numerical grid resolutions used in this study. $\Delta x$ and $\Delta z$ indicate the spacing of the uniform grid in $x$ and $z$ direction. In wall-normal $y$ direction, a non-uniform grid with spacing between $\Delta y_{min}$ and $\Delta y_{max}$ is used. All spacings are provided in outer units and in inner plus units, i.e. in units of $\delta_\tau$.
    The DNS-grid is used for direct numerical simulations, while LES simulations are carried out with the LES-grid.}
    \label{tab:resolution}
\end{table}

\section{Results and discussion}\label{sec:results}

In this section, we demonstrate that large-scale motions, LSMs, can be isolated from near-wall small-scale structures. First, we determine a reference value of the Smagorinsky constant $C_s$ so that the LES reproduces statistical properties of a resolved DNS. The original overfiltering technique by \cite{Hwang2010} is shown to not be able to filter out all small-scale structures without modifying properties of large scales. We however propose a modified overfiltering approach that ensures a physically correct mean profile and allows us to fully isolate LSMs from small-scale structures without modifying their properties. This demonstrates that LSMs are self-sustained in the ASBL. Finally, we describe details of the self-sustained process based on the dynamics of a single LSM periodically replicated in the horizontal plane.

\subsection{Reference case - LES reproducing DNS statistics}\label{subsec:refCase}

\begin{figure}
    \centering
    $(a)$
    \subfloat{{\includegraphics[width=0.45\textwidth]{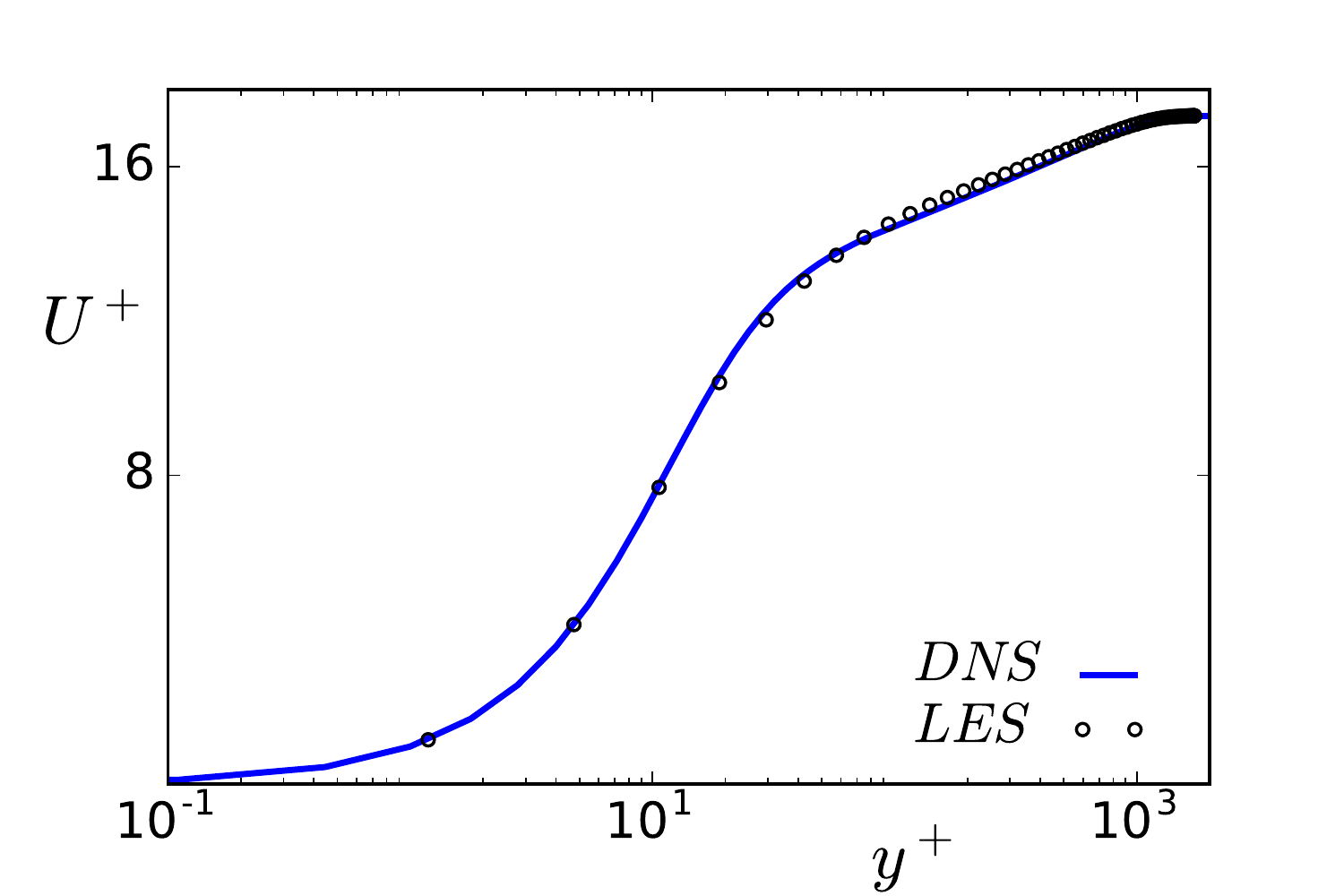} }}
    $(b)$
    \subfloat{{\includegraphics[width=0.45\textwidth]{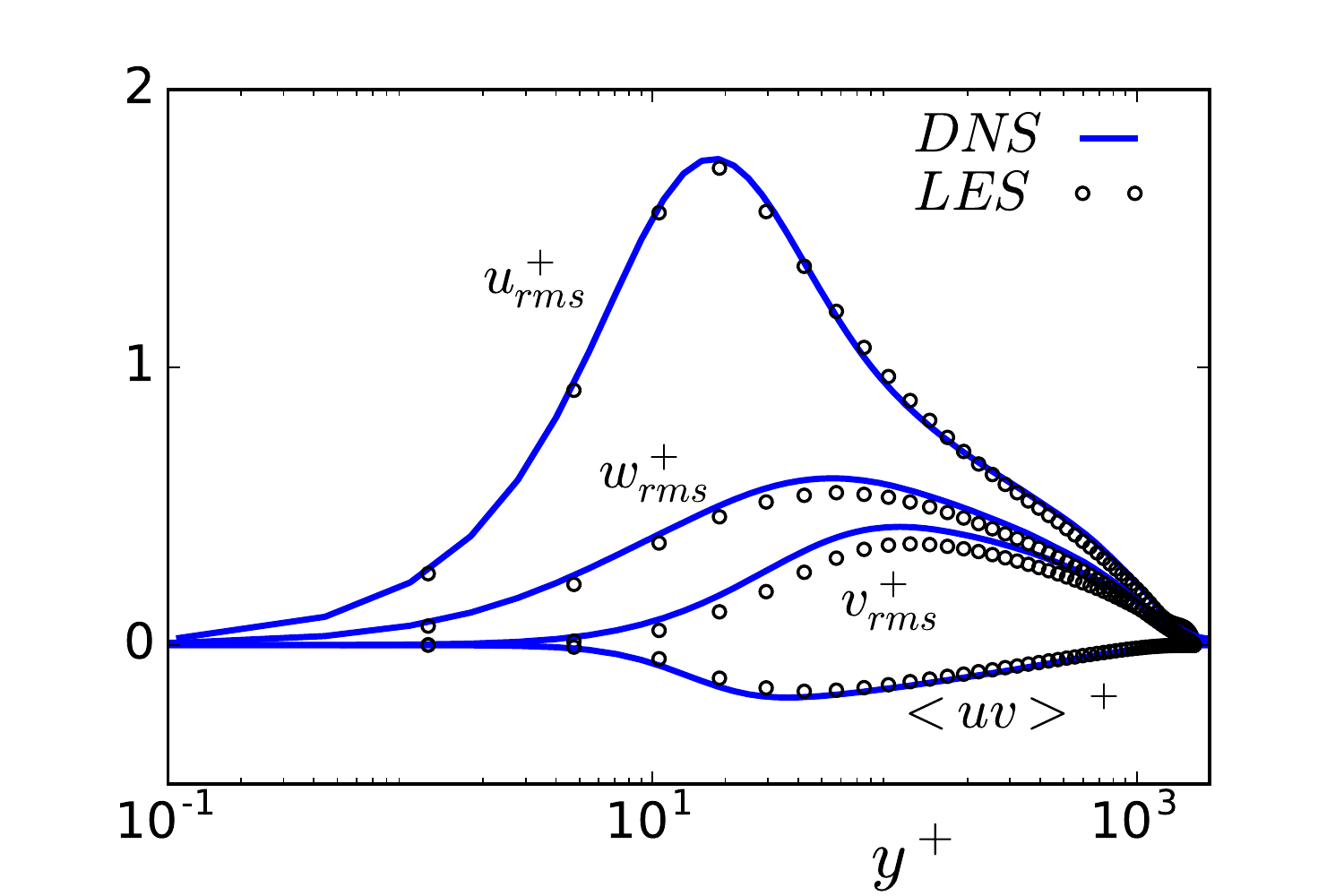} }}
    \caption{$(a)$ Mean streamwise velocity, and $(b)$ root mean squared (r.m.s.) velocity components as functions of the wall-normal coordinate $y$, for both the DNS (blue solid line) and the LES with the optimal value of $C_s=0.045$ (hollow circles).
    The mean and r.m.s. profiles from LES are in excellent agreement with the DNS. 
    We thus choose $C_s=0.045$ as reference value for LES reproducing DNS results.}
\label{fig:Cs_ref}
\end{figure}

We consider the flow at the Reynolds number $Re=300$ to obtain the reference value of the Smagorinsky constant $C_s$ at which the statistics of the LES best matches DNS results.
A DNS is performed in a domain of extension $L_x = 256,\ H = 150,\ L_z = 128$ discretised with $N_x = 256,\ N_y = 241,\ N_z = 256$ points (the RefDNS domain from table \ref{tab:domains}). 
We carry out several LES in a domain of extension $L_x = 256,\ H = 100,\ L_z = 128$ resolved with $N_x = 64,\ N_y = 61,\ N_z = 64$ points (the RefLES domain from table \ref{tab:domains}) and vary the value of the Smagorinsky constant $C_s$ until statistical properties of the LES solution match those of the DNS. For $C_s=0.045$ the first- and second-order statistics of the LES very well match DNS results, as shown in figure \ref{fig:Cs_ref}. 
In the rest of this article, $C_s=0.045$ is used as the reference value of Smagorisnky constant for LES simulations, performed in domains resolved with the LES-grid resolution from table \ref{tab:resolution}.

\subsection{Overfiltered LES using the original approach of Hwang and Cossu}\label{subsec:ODL_SSM}

Following the original overfiltering approach of \cite{Hwang2010}, we perform various LES for increasing values of the Smagorinsky constant $C_s$. The aim is to quench active small- and intermediate-scale structures by increasing $C_s$ beyond the reference value at which the LES reproduces DNS results. 
The simulations are carried out in a domain with $L_x = 512,\ H = 100,\ L_z = 256$ discretised with $N_x = 128,\ N_y = 61,\ N_z = 128$ collocation points (LESbox from table \ref{tab:domains}). We integrate the governing equations in time until statistical steady state is reached. 
For the reference $C_s=0.045$ the boundary layer thickness in steady state reaches $\delta_{99}=67.45$, which corresponds to a friction Reynolds number of $Re_\tau=1\,168$. In outer scale units, i.e. in units of $\delta_{99}$, the LESbox has a size of $(L_x \approx 8\delta_{99}, H \approx 1.5\delta_{99}, L_z \approx 4\delta_{99})$. The computational domain consequently accommodates several LSMs coexisting in the spanwise and in the streamwise directions.

\begin{figure}
    \centering
    $(a)$
    \includegraphics[width=0.35\textwidth]{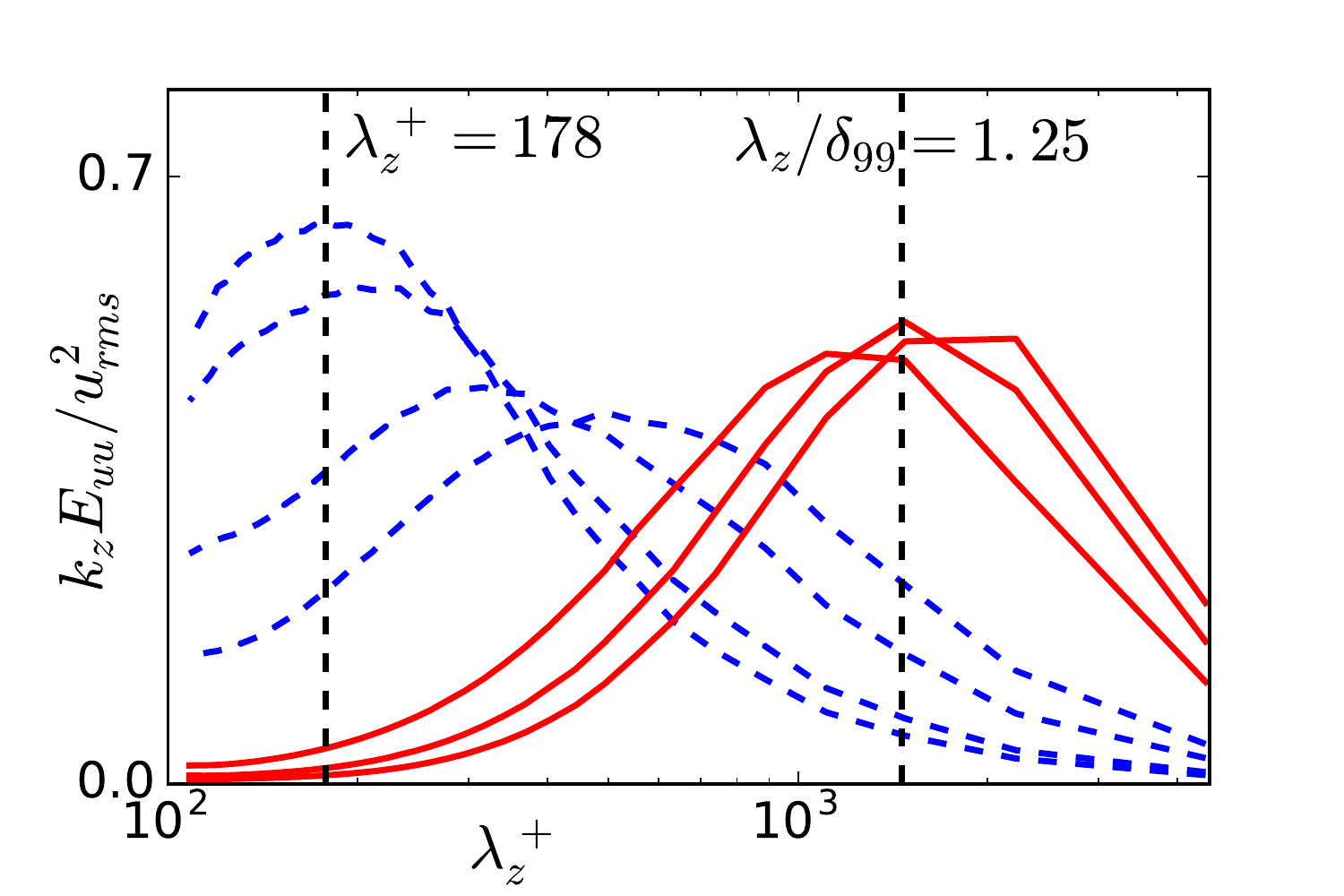}
    $(b)$
    \includegraphics[width=0.35\textwidth]{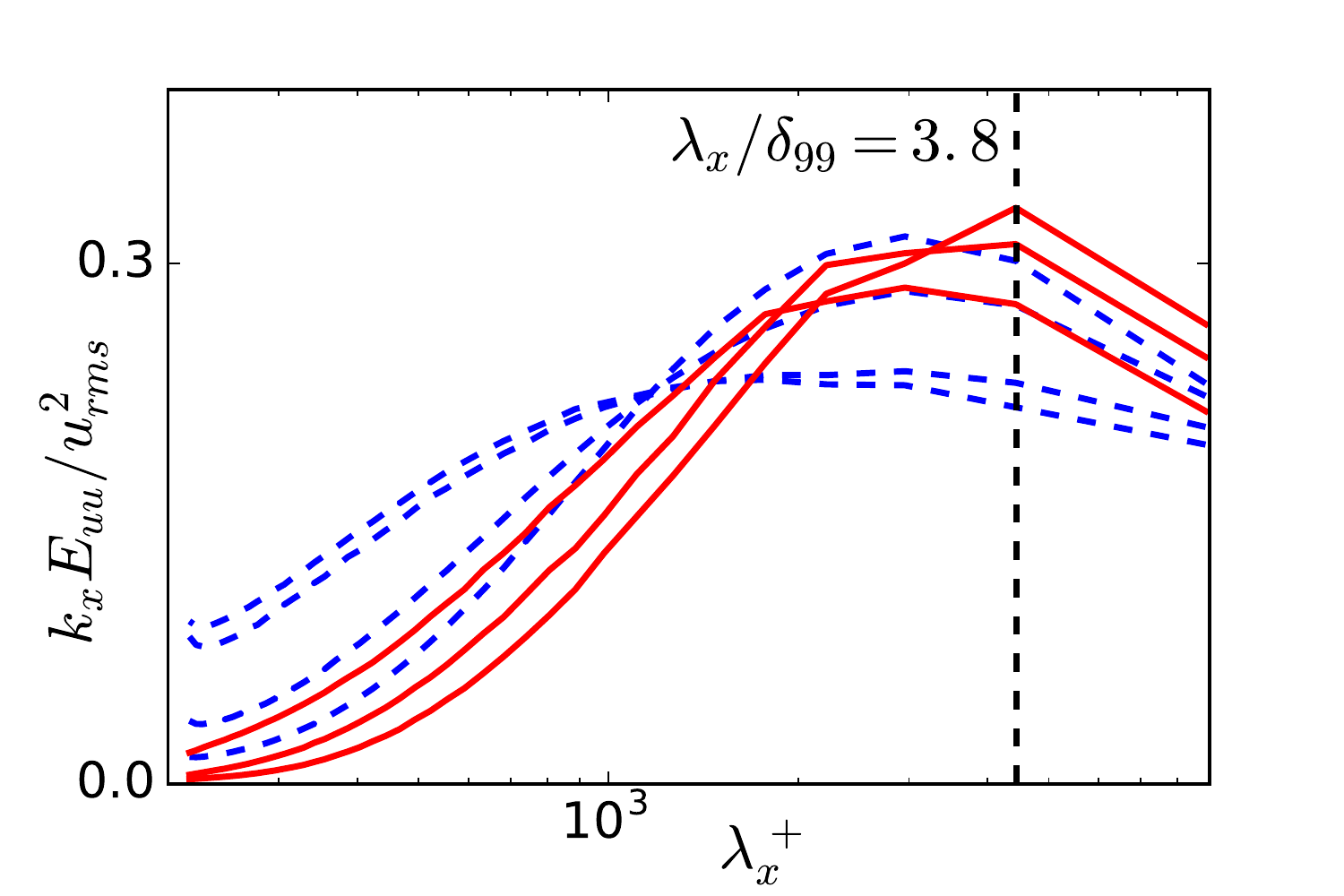}\\
    $(c)$
    \includegraphics[width=0.35\textwidth]{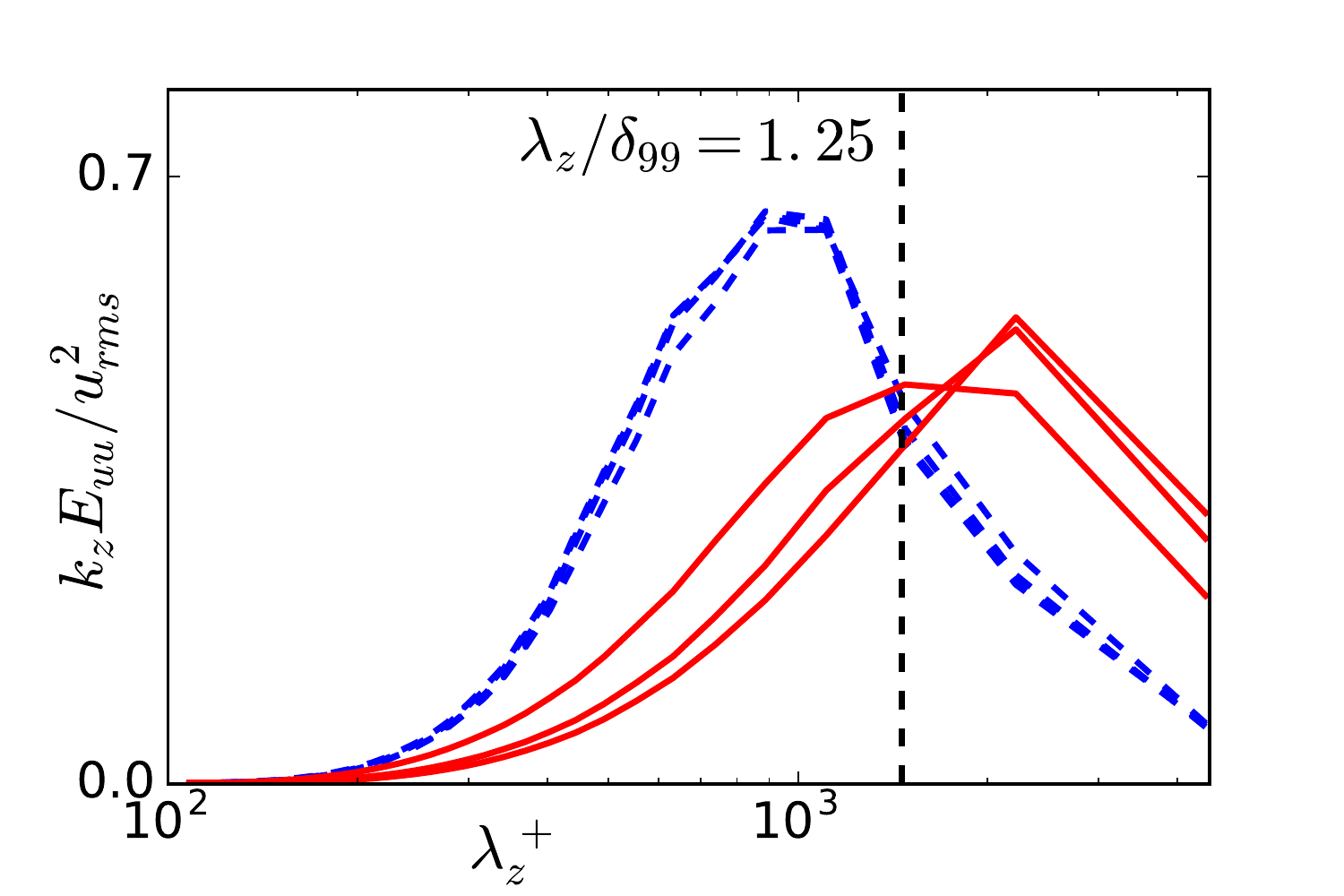}
    $(d)$
    \includegraphics[width=0.35\textwidth]{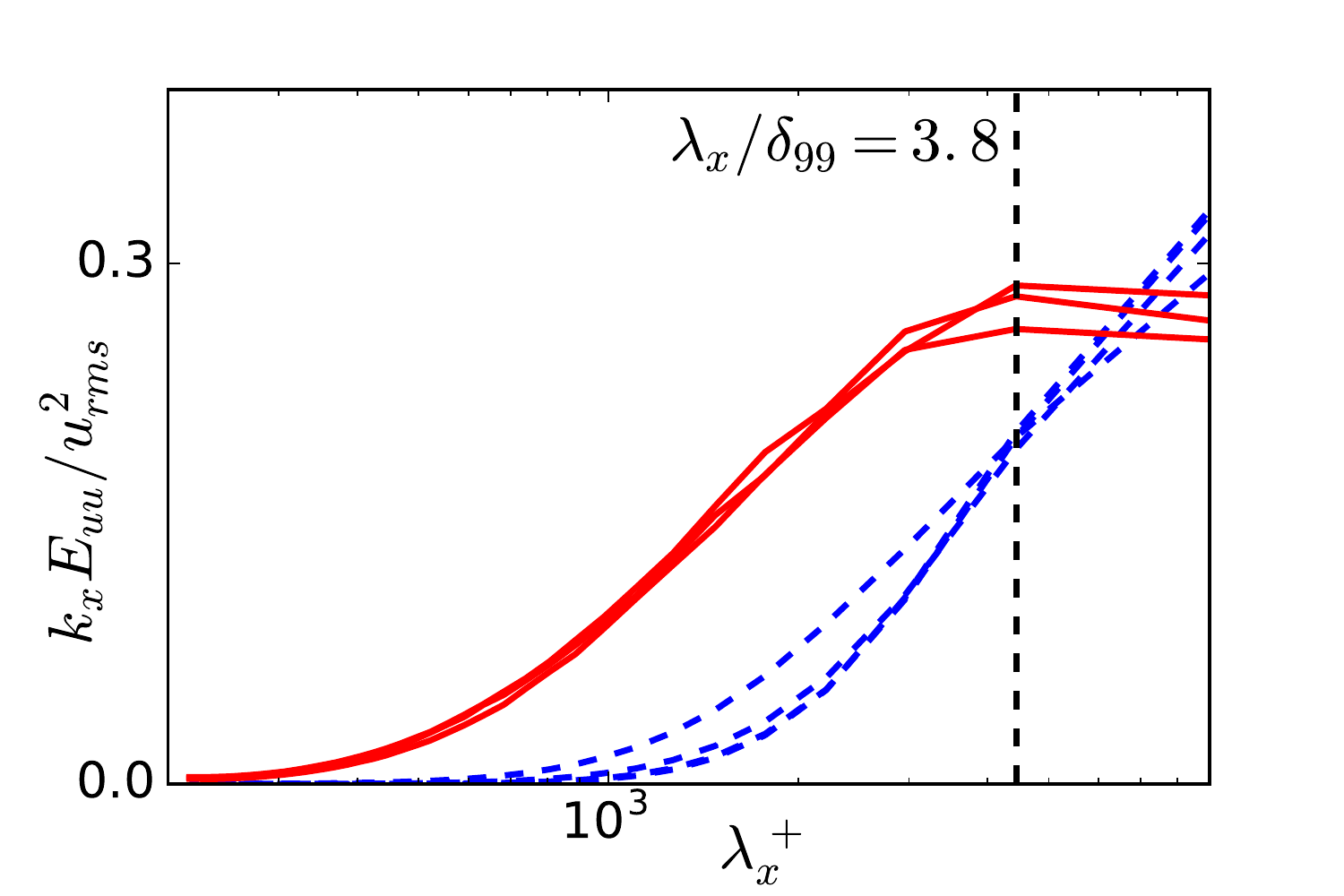}
    \caption{Spanwise premultiplied power spectra of the streamwise velocity (panels $a$ and $c$), and streamwise premultiplied power spectra of the streamwise velocity (panels $b$ and $d$) obtained by the static Smagorinsky model with two different values of $C_s$: the reference value $C_s = 0.045$ (panels $a$ and $b$); and a moderately increased value $C_s = 0.2$ (panels $e$ and $f$). The data are extracted in the inner region at $y^+ = \left[19, 30, 58, 94\right]$ (blue dashed lines) and in the outer region at $y/\delta_{99} = \left[0.3, 0.51, 0.74\right]$ (red solid lines).
    Relative to the reference case, at elevated $C_s$, the energy peak corresponding to the large-scale motions (represented by red solid lines) moves towards larger scales. The inner peak capturing small-scale near-wall structures (represented by blue dashed lines) is damped but intermediate-scale structures (peak at $y^+ \approx 1000$ in the near-wall region) remain active. Consequently, there is no strength of overfiltering, where all small- and intermediate-scale structures are quenched without deteriorating the LSMs. 
    }
\label{fig:ODL_SSM}
\end{figure}

We identify characteristic turbulent structures by computing both streamwise and spanwise premultiplied streamwise velocity power spectra for wall-parallel planes at multiple distances from the wall, as shown in \reffig{ODL_SSM}. For distances in the near wall region ($y^+<100$) a peak associated with near-wall small-scale structures is observed, while further from the wall ($y\ge 0.3 \delta_{99}$) a peak characterising LSMs emerges. We analyse the effect of increased $C_s$ on the turbulent structures by observing modifications and shifts of the energy peaks characterising small-scale structures near the wall and LSMs, respectively. 
For the reference $C_s=0.045$, the spanwise premultiplied spectra (panel $a$) show the near-wall peak at $\lambda_z^+\approx 178$, while the peak characterising LSMs in the outer region is located at $\lambda_z=1.25\,\delta_{99}$.
In streamwise direction (panel $b$), the near-wall peak is located at $\lambda_x^+ \approx 1750$, and the outer peak at $\lambda_x \approx 3.8\,\delta_{99}$.

The aim of the overfiltering approach proposed by \cite{Hwang2010} is to completely quench spectral peaks associated with small- and intermediate-scale active motions for values of $C_s$ that do not deteriorate the large-scale features of the flow.
However, while the approach was successful in confined flows, in the boundary layer considered here, the quantitative statistical properties of the large-scale motions are significantly affected by the filtering for values of $C_s$ below the value required to completely quench small-scale structures.
This is demonstrated in figure \ref{fig:ODL_SSM}($c$ and $d$) where the energy spectra corresponding to the Smagorinsky constant $C_s=0.2$ are reported.
At this moderate $C_s=0.2$, the near-wall peak at spanwise wavelength $\lambda_z^+\approx 178$ has been successfully damped but intermediate-scale structures remain active, as evidenced by a clear peak at $y^+ \approx 1000$. While the moderate value of $C_s$ is apparently not sufficient to quench all small-scale structures, the filtering already significantly distorts large-scale motions as evidenced by the spanwise outer peak shifting by approximately $50\%$ relative to the reference case, to $\lambda_z \approx 1.9 \delta_{99}$. Further evidence that already at $C_s=0.2$ LSMs are significantly distorted by the overfiltering is given by the fact that the boundary layer thickness $\delta_{99}$ significantly grows and fails to saturate before interacting with the non-physical upper wall of the computational domain.  
In ASBL there is therefore no value of $C_s$ at which small-scale structures are quenched without significantly deteriorating the LSMs. Consequently, isolating LSMs and investigating if LSMs are self-sustained requires a modified overfiltering approach. 

\subsection{Modification of the original overfiltering approach}\label{subsec:modification}

As discussed in the previous section, the established overfiltering approach fails to isolate large-scale motions from smaller-scale structures in the ASBL.
The large-scale motions are strongly affected by the filtering at the value of $C_s$ that is required to damp the strong small-scale structures in the near-wall region.
This problem might stem from the fact that compared to near-wall structures large-scale motions in the ASBL are weaker than in turbulent channels and Couette flow, where the established overfiltering approach was successful. Possibly due to the difference in relative strength, smaller-scale motions are alive at the values of $C_s$ that significantly affect the LSMs.
If $C_s$ is sufficiently increased to quench all smaller-scale motions, the flow is thus left with significantly deteriorated LSMs having spatial scales which are significantly larger than those observed in the reference case representing a real physical flow.  

The deterioration of LSMs is associated with the non-physical growth of the boundary layer thickness representing an incorrect mean velocity profile. These non-physical modifications of large-scale mean flow properties reflect the errors introduced by overfiltering with a subgrid model that can only capture some of the physics of turbulence at scale smaller than the filter width.  
Since we have information on the physically correct mean profile based on DNS and the reference LES, the key idea is to use this additional information to correct for the errors introduced by the subgrid model in overfiltered simulations. We thus propose a modified overfiltering procedure where the known and physically correct mean profile is imposed. 

\begin{figure}
    \centering
    $(a)$
    \includegraphics[width=0.41\textwidth]{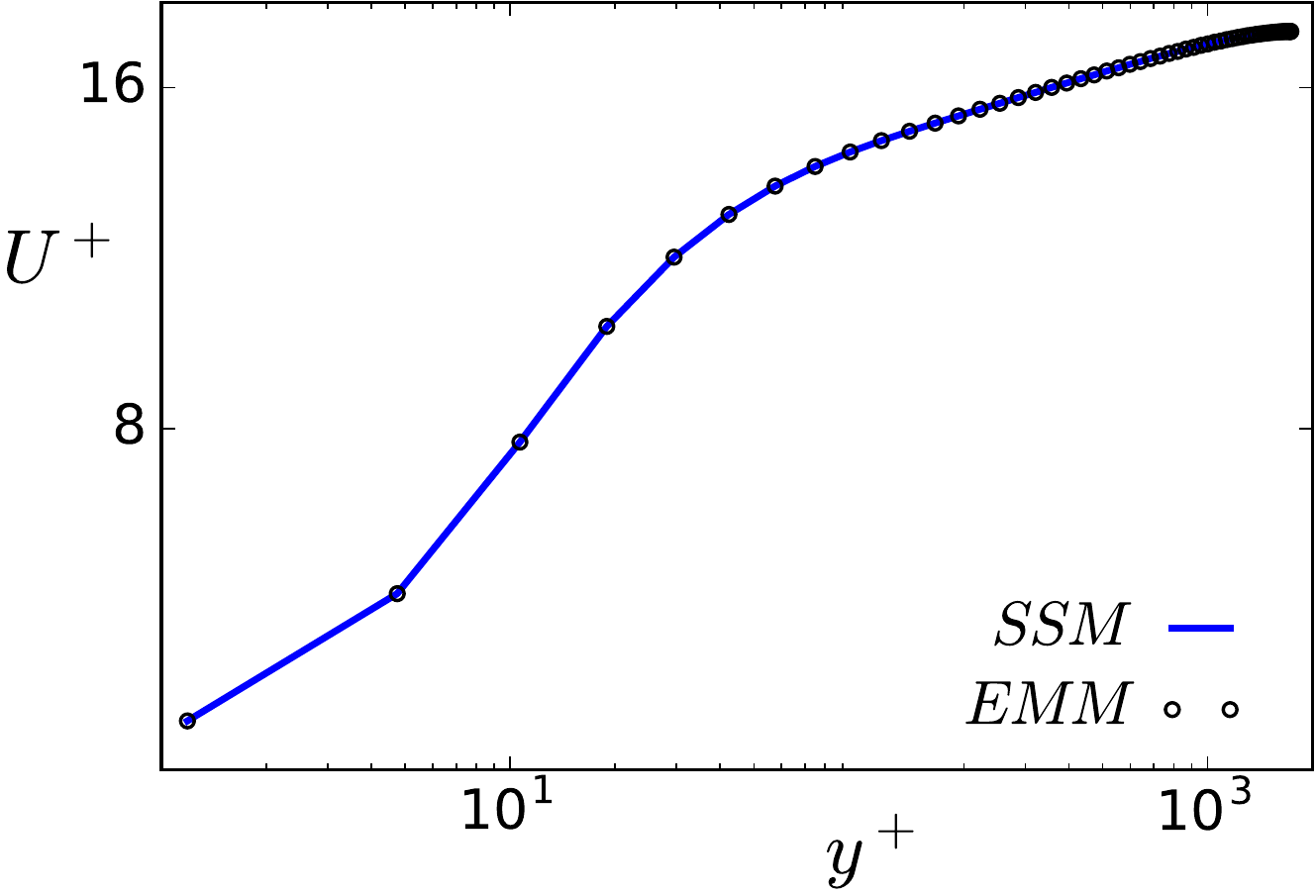}
    $(b)$
    \includegraphics[width=0.38\textwidth]{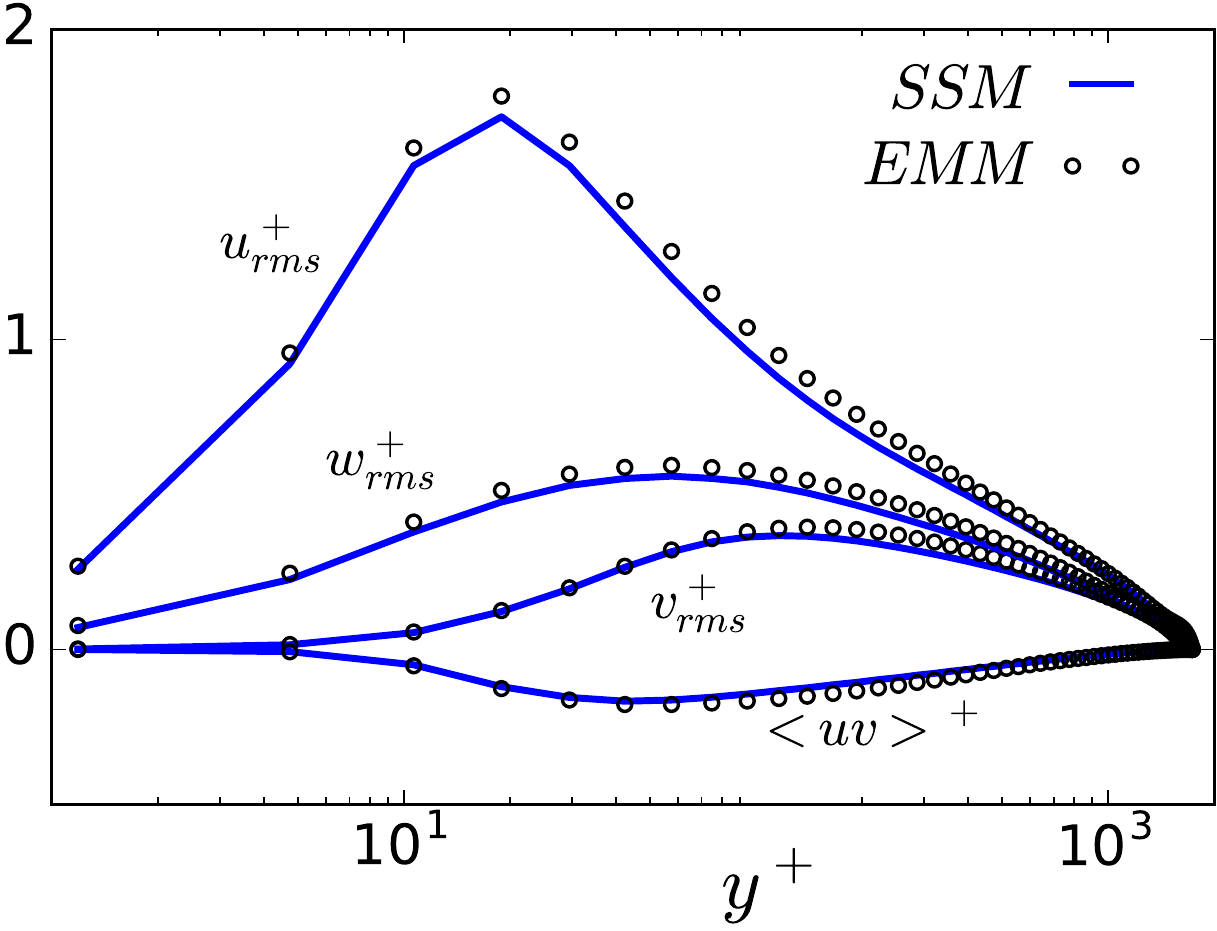}
    \caption{$(a)$ Mean velocity profile, and $(b)$ root mean squared (r.m.s.) velocity fluctuations as functions of the wall-normal coordinate $y$, at the reference value of $Cs=0.045$ for the static Smagorinsky model (SSM) and the enforced mean velocity model (EMM).
    Results from the EMM (hollow circles) match those of the SSM (blue solid lines). Consequently, the EMM faithfully reproduces correct velocity fluctuations. }
\label{fig:EMM_ref}
\end{figure}

The mean profile is given by the $(0,0)$ Fourier harmonic of the streamwise velocity. The mode is given by  $\tilde{u}_{0,0}(y,t) = \langle \overline{u}(x,y,z,t)\rangle_{x,z}$, where the angle brackets indicate the spatial average in stream- and spanwise direction. Technically, we enforce the correct $(0,0)$ mode in each time step of the LES, $\tilde{u}_{0,0}(y,t) \equiv U(y)$, where $U(y)$ is the known mean profile. All other harmonics are computed as usual.
The modified modelling approach, using the static Smagorinsky model \emph{with} enforced mean velocity (from here on termed EMM), reproduces turbulent fluctuations computed from the established pure static Smagorinsky model without enforced mean (from here on termed SSM), when the reference Smagorinsky constant $C_s=0.045$ is used. This is confirmed by matching r.m.s. profiles shown in figure \ref{fig:EMM_ref}.

\subsection{Overfiltered LES with enforced mean velocity profile}\label{subsec:ODL_EMM}

\begin{figure}
    \centering
    $(a)$
    \includegraphics[width=0.35\textwidth]{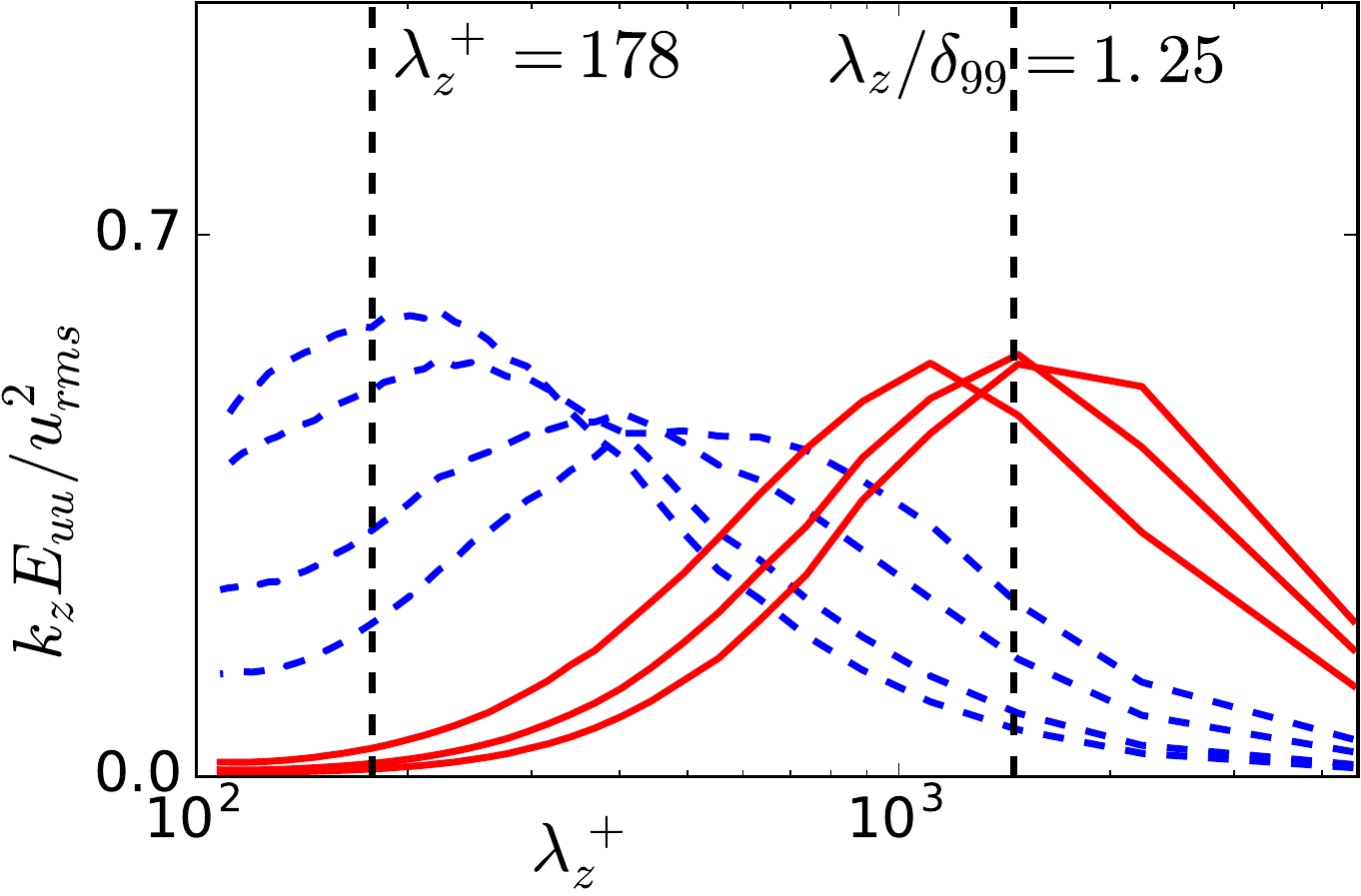}
    $(b)$
    \includegraphics[width=0.35\textwidth]{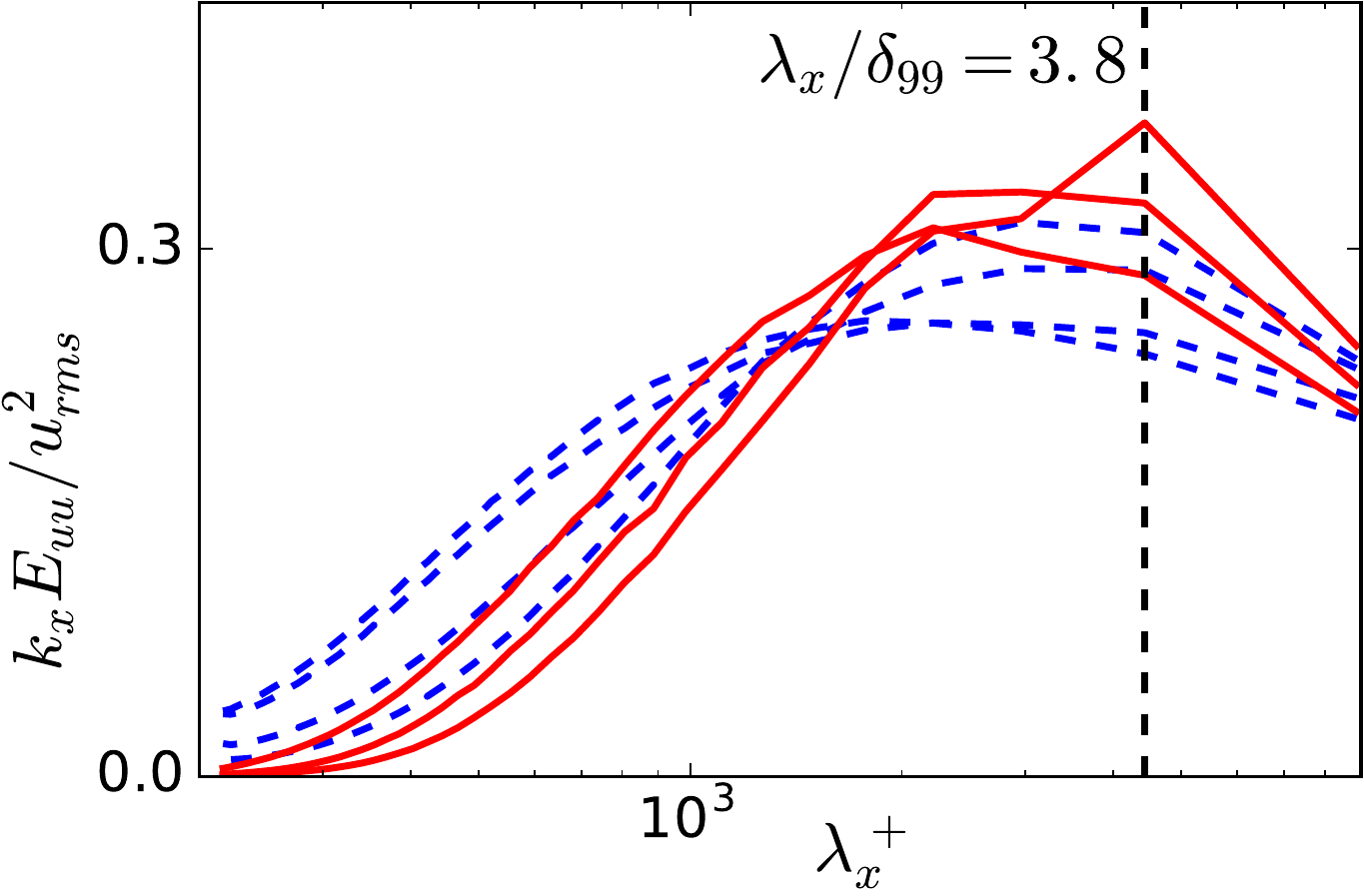} \\
    $(c)$
    \includegraphics[width=0.35\textwidth]{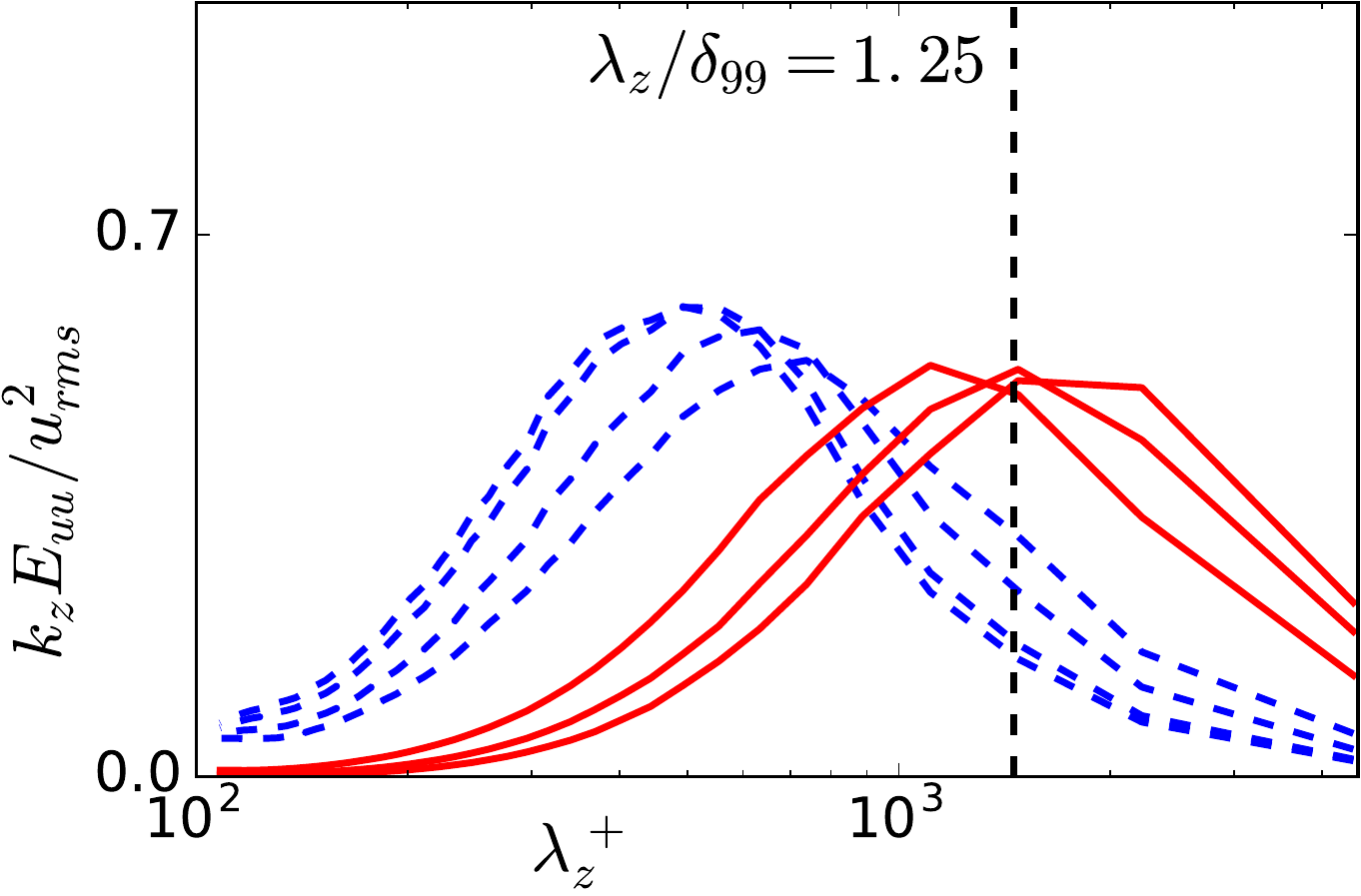}
    $(d)$
    \includegraphics[width=0.35\textwidth]{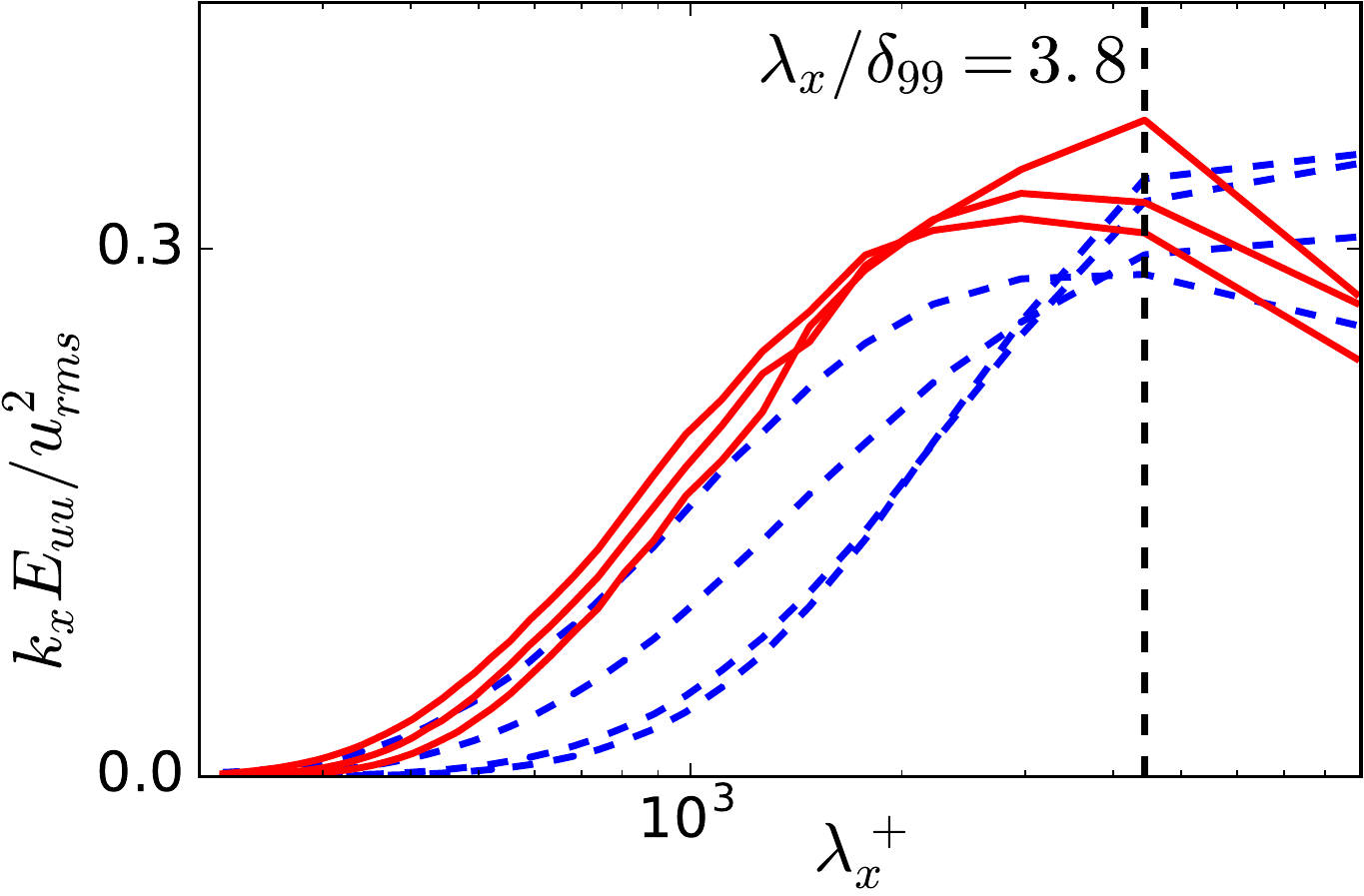}\\
    $(e)$
    \includegraphics[width=0.35\textwidth]{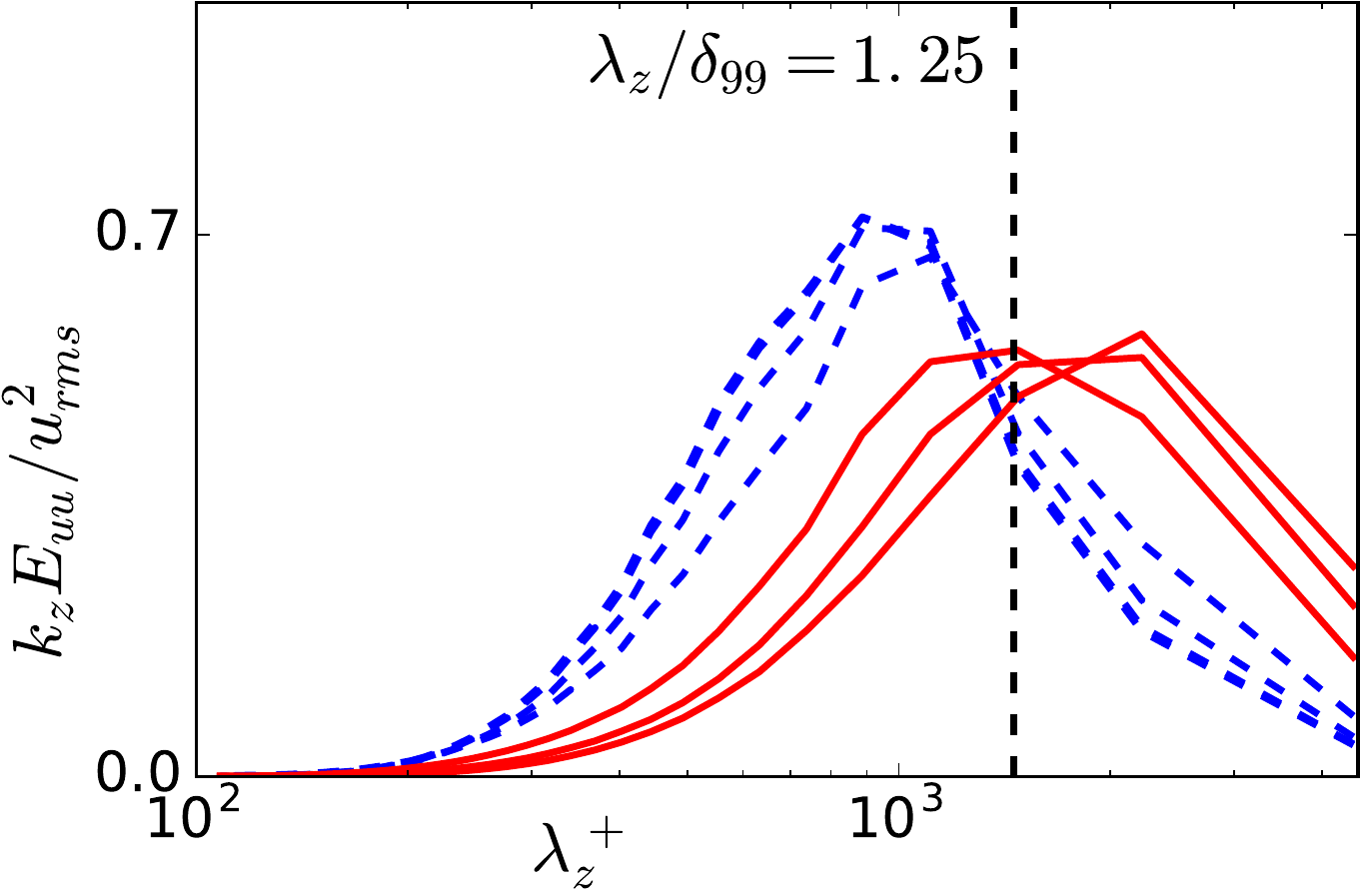}
    $(f)$
    \includegraphics[width=0.35\textwidth]{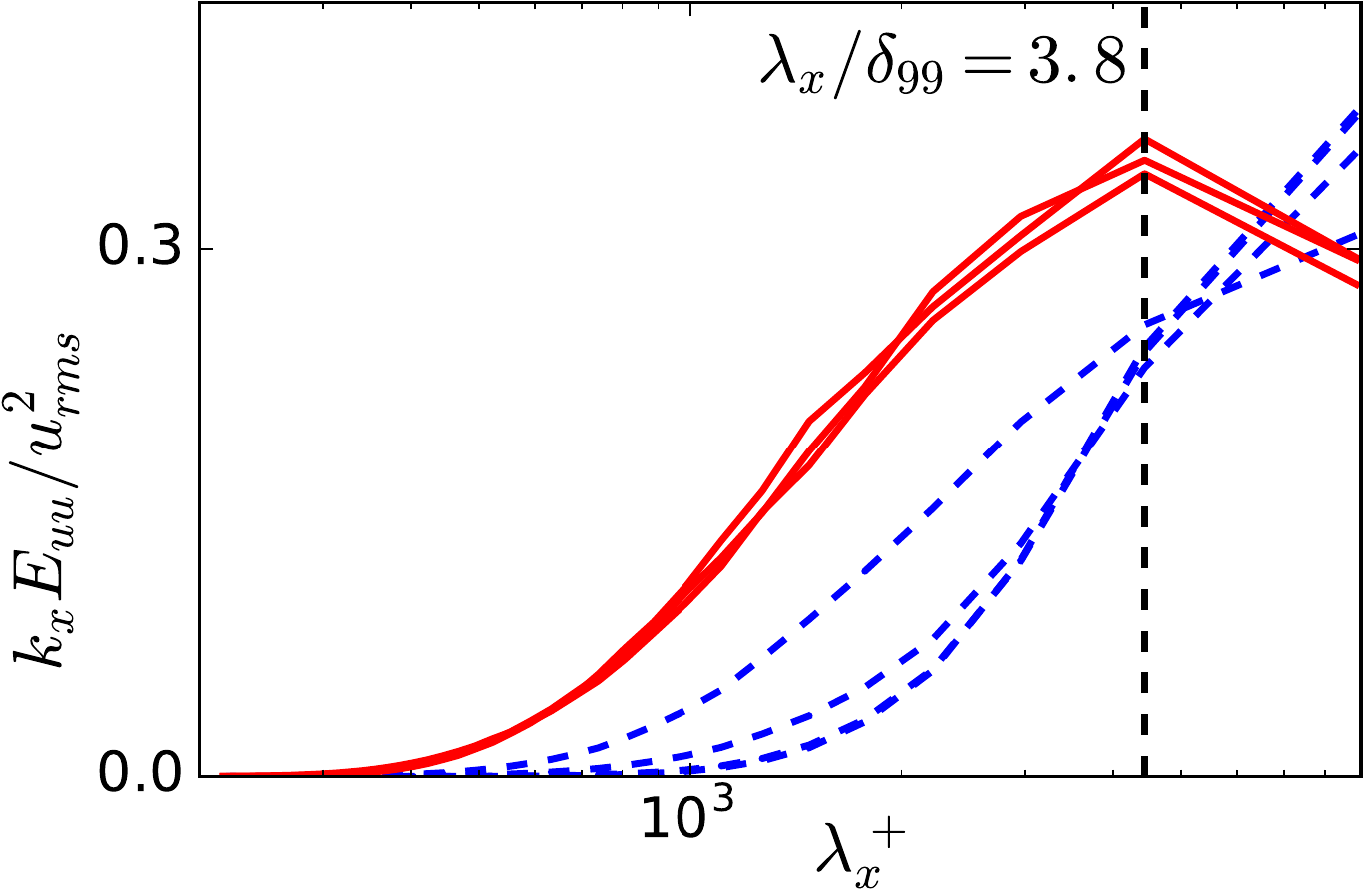}\\
    $(g)$
    \includegraphics[width=0.35\textwidth]{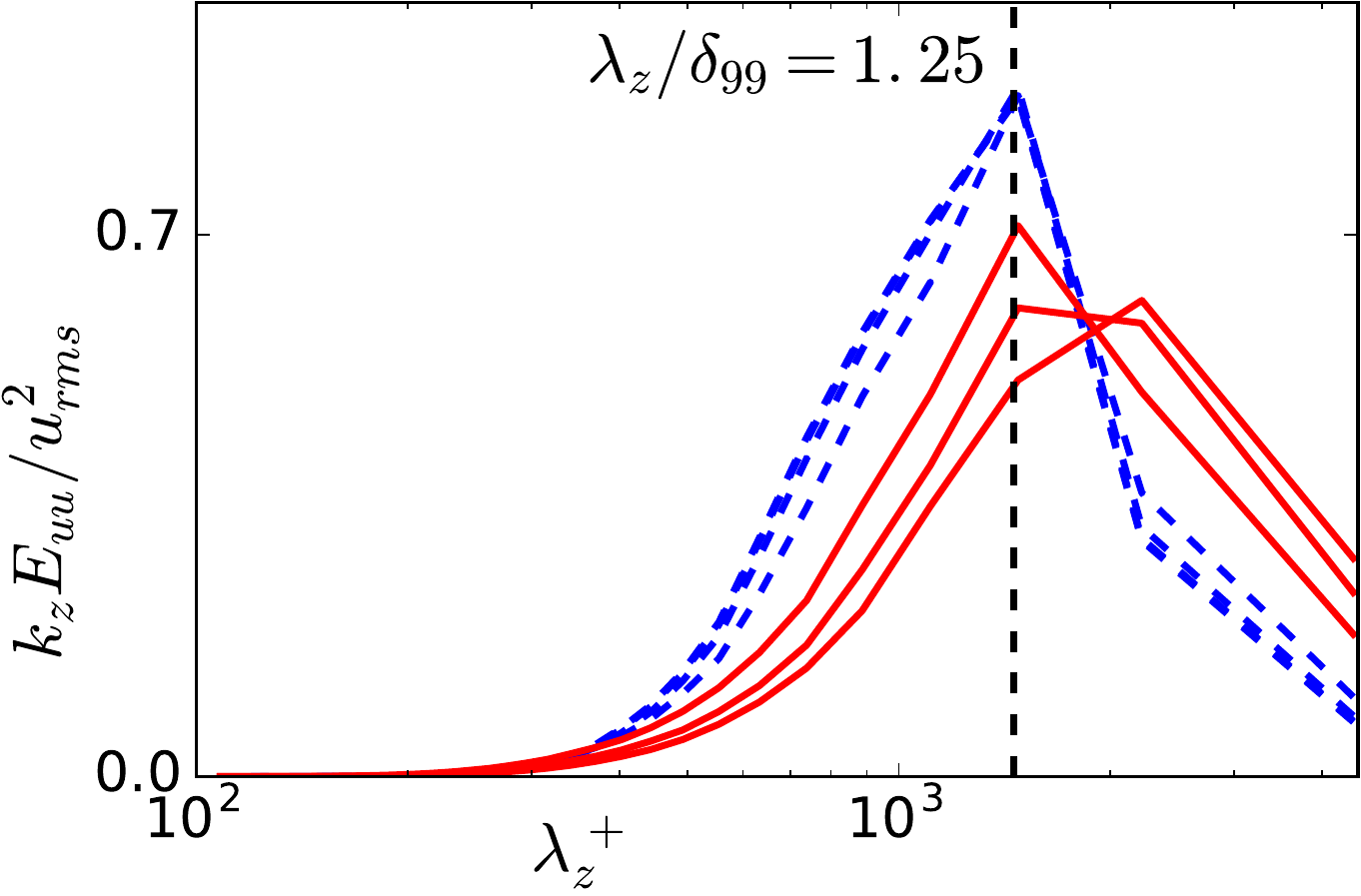}
    $(h)$
    \includegraphics[width=0.35\textwidth]{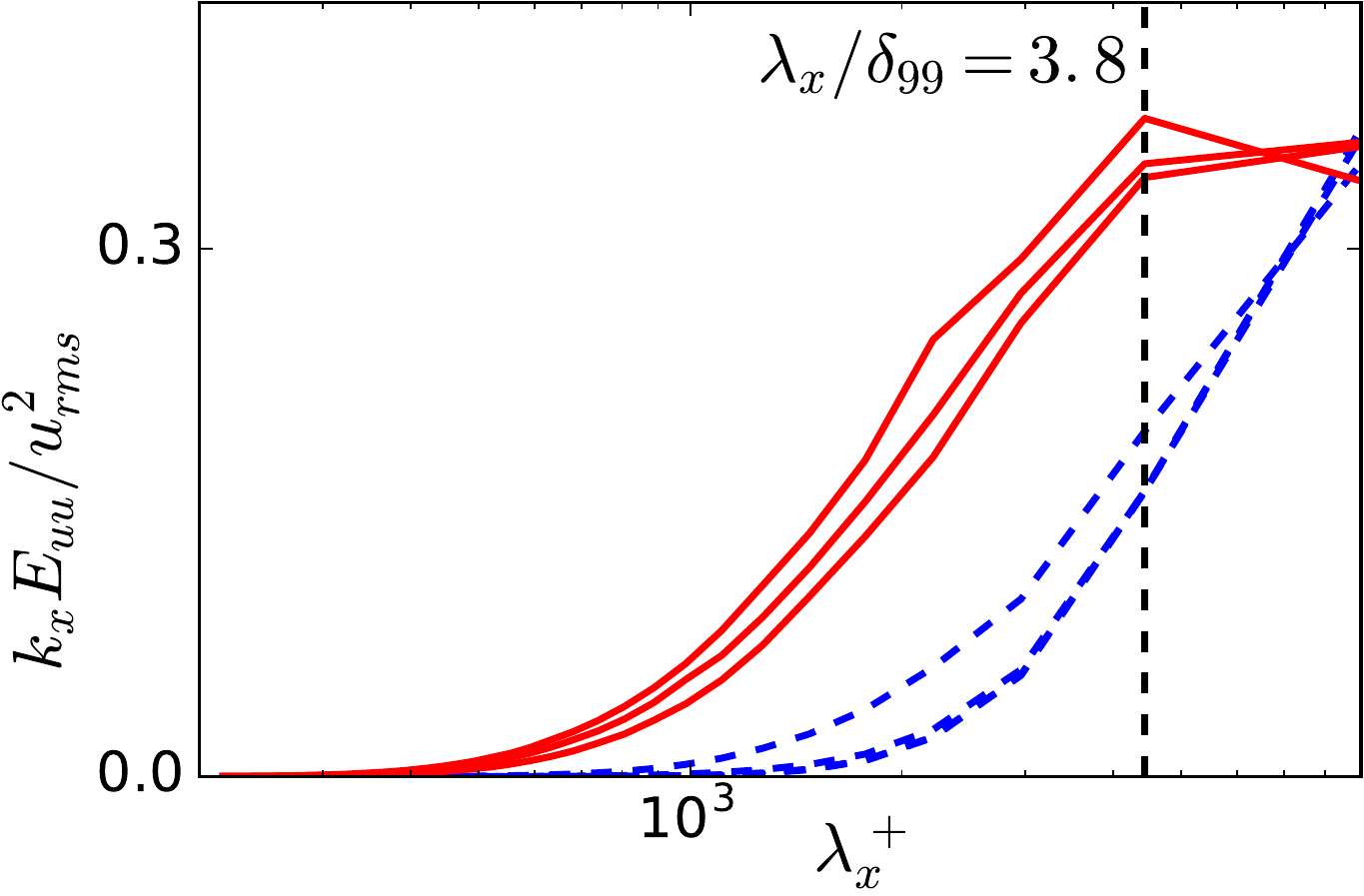}\\
    \caption{Spanwise premultiplied power spectra of the streamwise velocity (panels $a$, $c$, $e$ and $g$), and streamwise premultiplied power spectra of the streamwise velocity (panels $b$, $d$, $f$ and $h$) obtained by the static Smagorinsky model with enforced mean velocity profile, EMM. The value of $C_s$ is increased from top to bottom: the reference value $C_s = 0.045$ (panels $a$ and $b$); $C_s = 0.1$ (panels $c$ and $d$); $C_s = 0.2$ (panels $e$ and $f$); and $C_s = 0.3$ (panels $g$ and $h$). As in figure \ref{fig:ODL_SSM} the data is shown for the inner near-wall region at $y^+ = \left[19, 30, 58, 94\right]$ (blue dashed lines) and for the outer region at $y/\delta_{99} = \left[0.3, 0.51, 0.74\right]$ (red solid lines).
    When $C_s$ is increased from the reference value, the energy peaks corresponding to the small-scale structures in the near-wall region (represented by blue dashed lines) shift towards larger scales while the energy peaks related to LSMs (represented by red solid lines) remain at their location. At $C_s=0.3$ (panels $g$ and $h$), all small and intermediate scales are damped while LSMs survive unchanged. 
    Thus, the modified filtering approach successfully isolates LSMs from smaller-scale structures suggesting LSMs are self-sustained. 
    }
\label{fig:ODL_EMM_bigBox}
\end{figure}

Overfiltered large-eddy simulations are repeated using the modified approach preserving the turbulent mean flow to determine if it is possible to isolate LSMs from the dynamics of the near-wall small-scale structures.
The simulations are carried out in the same numerical domain $(L_x=512,\ H = 100,\ L_z = 256)$ with the same grid resolution (see table \ref{tab:resolution}) and the same Reynolds number ($Re=300$) as in section \ref{subsec:ODL_SSM}.
It is found that, as before, the reference value of Smagorinsky constant $C_s=0.045$ is the one best reproducing the results of the DNS.

As previously, premultiplied power spectra in both the near-wall and the outer region are used to quantify the effect of the filtering on small-scale structures near the wall and LSMs. As shown in figure \ref{fig:ODL_EMM_bigBox}, for the reference $C_s=0.045$ (panels $a$ and $b$) the peaks in the premultiplied spectra remain at their usual locations corresponding 
to the small-scale structures in the buffer layer ($\lambda_z^+ = 178$, $\lambda_z^+ = 1750$) 
and to the LSMs ($\lambda_z=1.25\, \delta_{99}$, $\lambda_z = 3.8\, \delta_{99}$).
When increasing $C_s$, the LSM peaks in the premultiplied spectra remain essentially unchanged while all peaks corresponding to smaller-scale structures are progressively quenched while shifting towards larger scales. 
At $C_s=0.3$ (panels $g$ and $h$), the near-wall peak at a spanwise wavelength of $\lambda_z \approx 178$ has been successfully damped and there is also no peak at intermediate scales indicating that at $C_s=0.3$ structures with spatial scales characteristic of buffer-layer and log-layer structures have been completely quenched. The outer peak (red lines), on the contrary, remains essentially unchanged both in spanwise and streamwise direction. We thus identified a value of the Smagorisnky constant at which all smaller-scale structures are successfully filtered out without distorting the LSMs. Consequently, the LSMs are successfully isolated from smaller-scale structures and appear to be self-sustained.

\begin{figure}
    \centering
    $(a)$
    \includegraphics[width=0.3\textwidth]{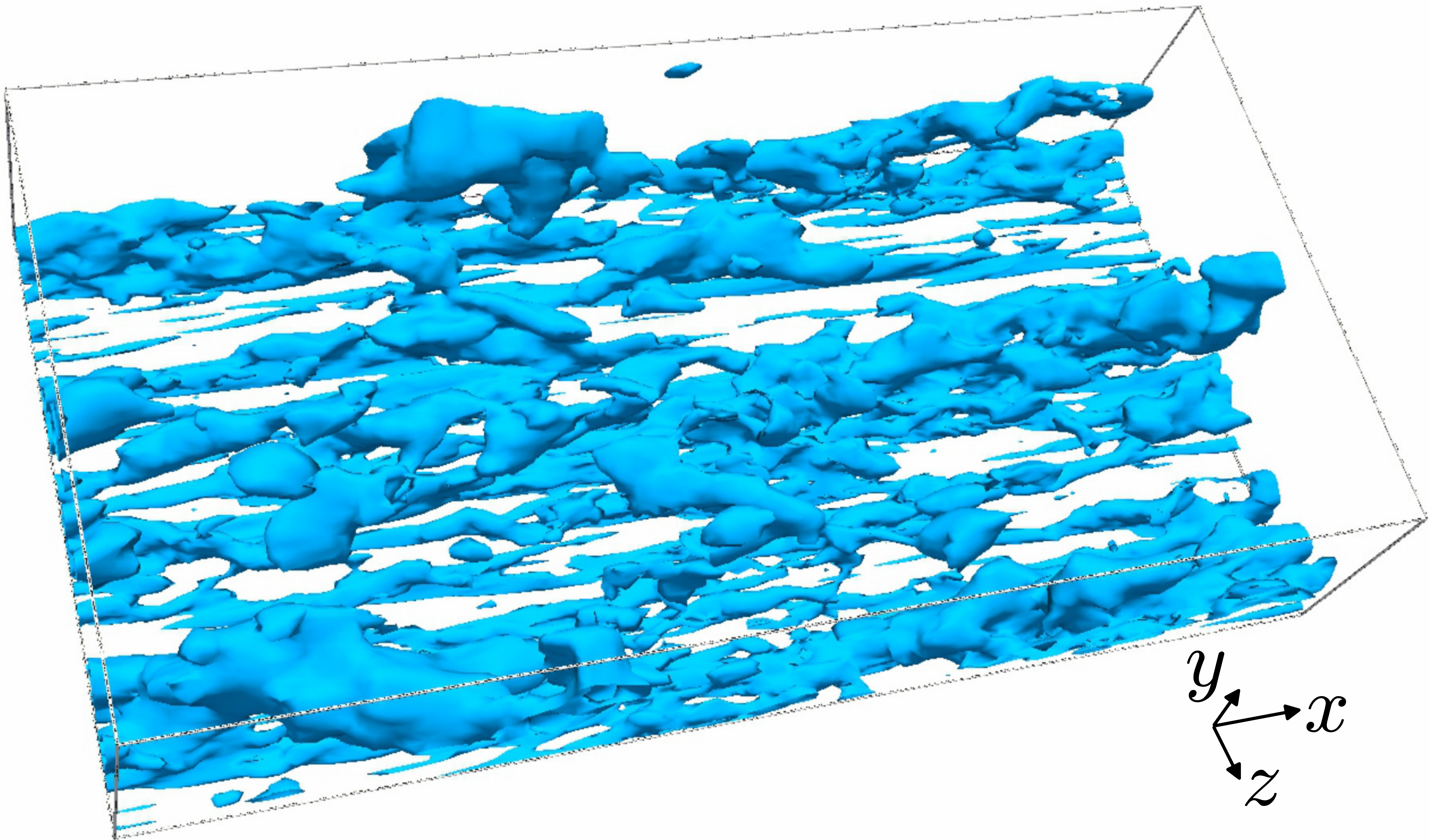}\\
    $(b)$
    \includegraphics[width=0.3\textwidth]{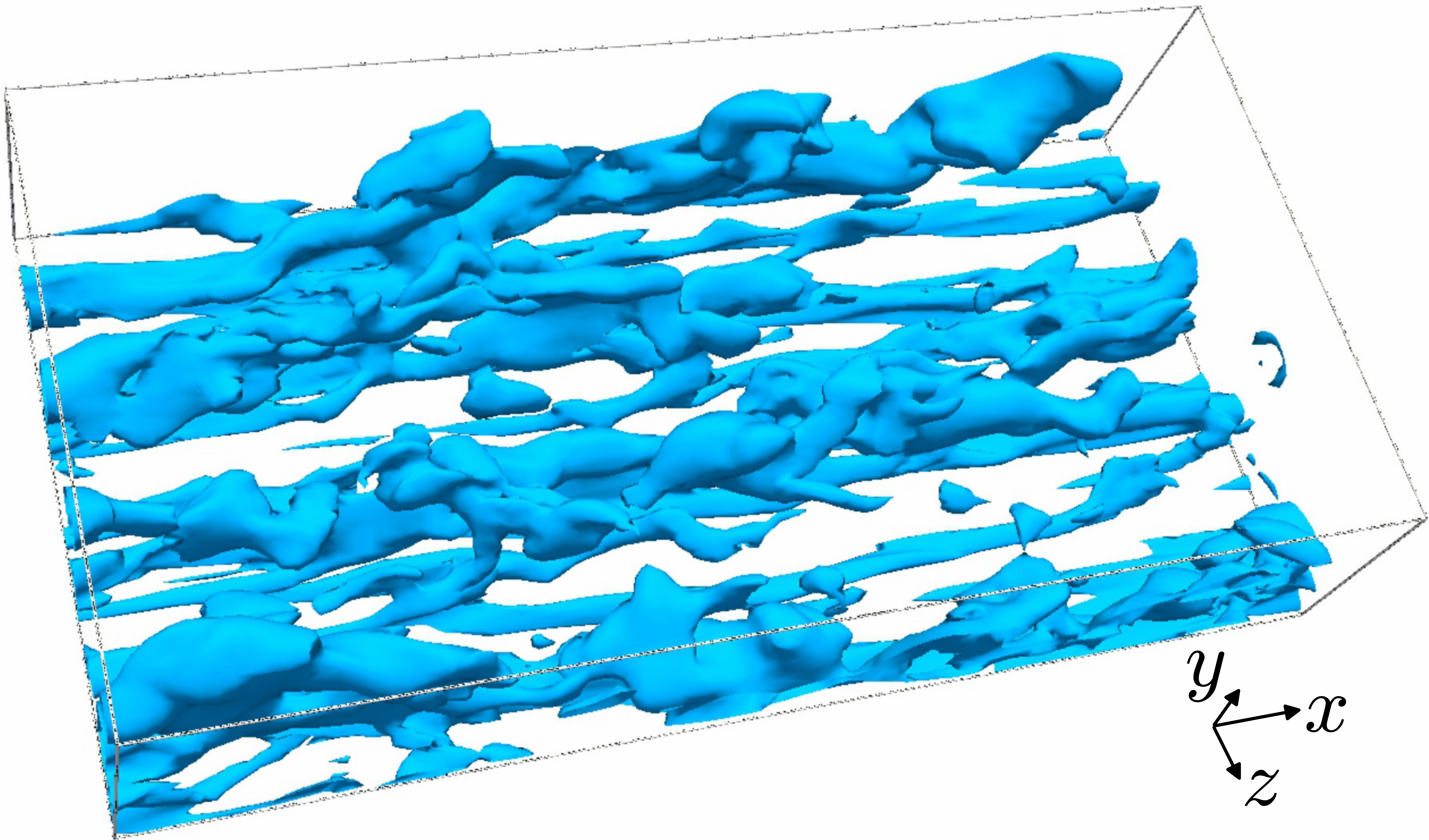}\\
    $(c)$
    \includegraphics[width=0.3\textwidth]{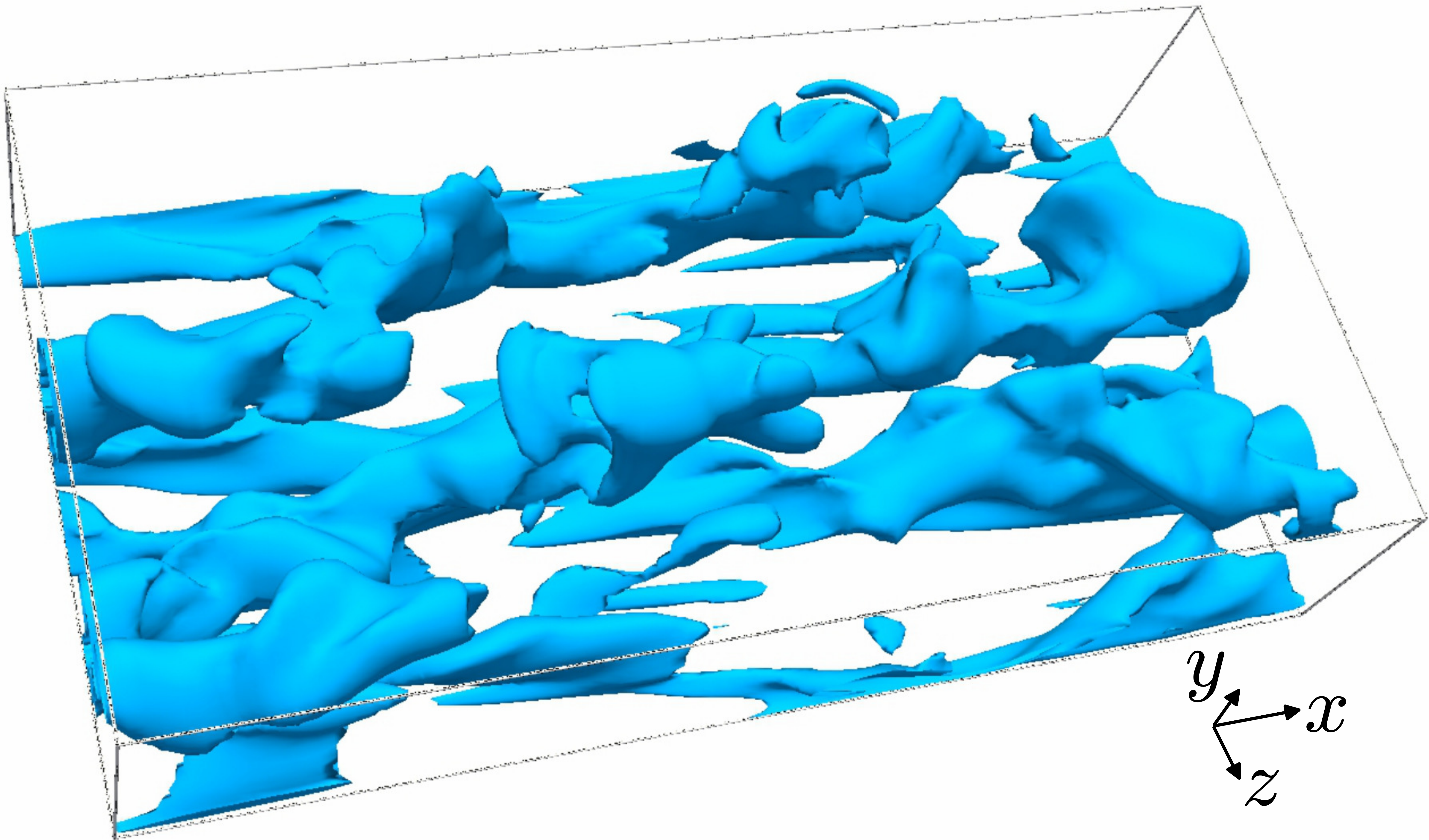}\\
    $(d)$
    \includegraphics[width=0.3\textwidth]{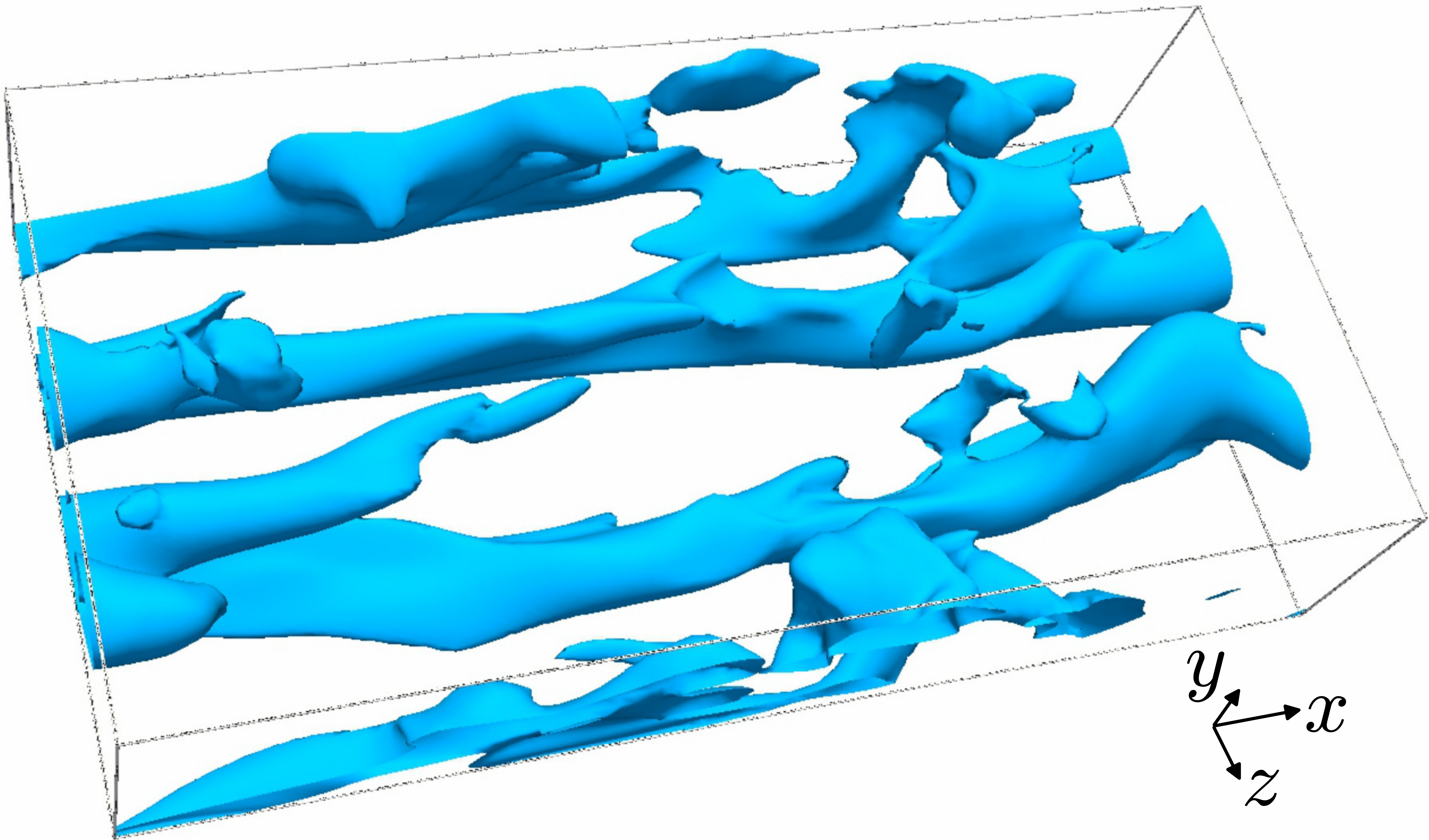}
    \caption{Instantaneous flow fields for increasing $C_s$: $(a)$ $C_s=0.045$ (the reference case), $(b)$ $C_s=0.1$, $(c)$ $C_s=0.2$, and $(d)$ $C_s=0.3$. Isosurfaces ($u^+=-0.52$) of the streamwise velocity deviation from the mean flow, i.e. low-speed streaks, are visualised.
    By increasing $C_s$, the velocity field of the LSMs become smooth, indicating that small-scale structures are damped out.
    }
\label{fig:filtering_fields}
\end{figure}
\begin{figure}
    \centering
    $(a)$
    \includegraphics[width=0.7\textwidth]{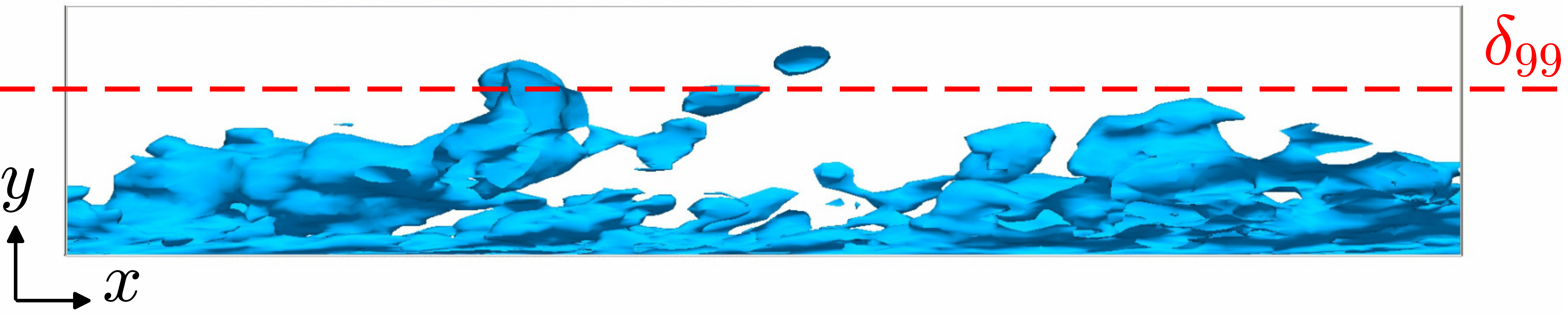}\\
    $(b)$
    \includegraphics[width=0.7\textwidth]{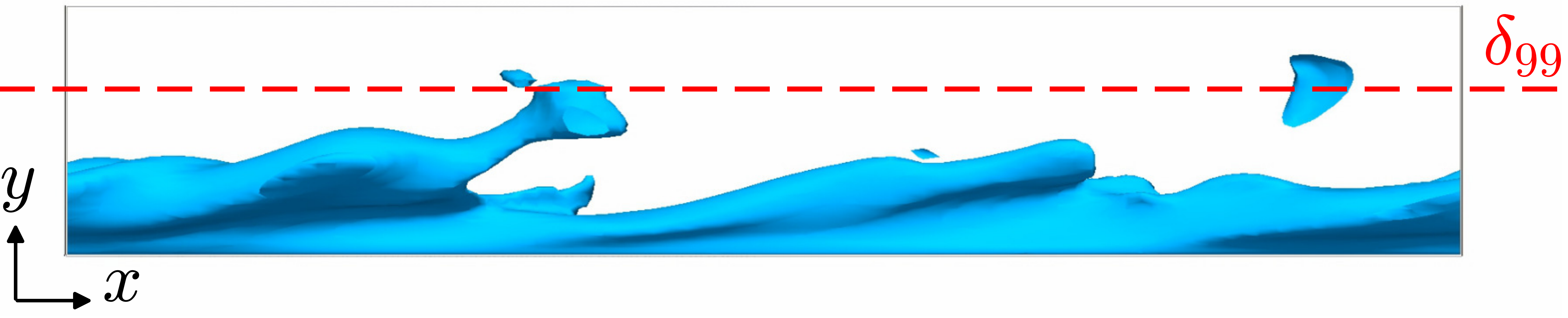}
    \caption{Lateral view of the instantaneous flow fields at $(a)$ $C_s=0.045$ (the reference case) and $(b)$ $C_s=0.3$. Isosurfaces ($u^+=-0.52$) of the streamwise velocity deviation from the mean flow are visualised, as in figure \ref{fig:filtering_fields}. The location of the turbulent boundary layer thickness $\delta_{99}$ is indicated by the dashed red line.
    The spatial structure of the isolated LSMs (panel $b$) resembles the spatial structure of the LSMs in the reference LES (panel $a$).
    }
\label{fig:filtering_fields_side}
\end{figure}

The effect of the overfiltering on the spatial characteristics of the flow structures is demonstrated in figure~\ref{fig:filtering_fields}, where flow snapshots for increasing values of $C_s$ are visualised.
As expected based on the power spectra, increasing $C_s$ removes the small-scale motions from the filtered flow while the characteristics of LSMs is preserved. For $C_s=0.3$, the streaks of LSMs have become smooth indicating that they are isolated from any smaller-scale velocity fluctuations.

Further evidence that the filtered LSMs are not distorted but resemble the spatial structure of the LSMs in the reference simulation is given in figure 
\ref{fig:filtering_fields_side}. Visualisations of low-speed streaks are compared between snapshots from the overfiltered simulation at $C_s=0.3$ and the reference case with $C_s=0.045$. The lateral view reveals that, indeed, the overfiltering does not change the scales of the LSMs. Specifically, the wall-normal scale on the order of the boundary layer thickness is preserved. 
Likewise, the ramp structure of LSMs, that is a characteristic of large-scale motions \citep{Hommema2003,Dennis2011,Rawat2015a}, is observed both in the reference simulation (panel $a$) and the overfiltered simulation (panel $b$).
The unchanged scales and preserved ramp structure further confirms that the overfiltered velocity field indeed represents genuine, undistorted and thus physically correct LSMs.

Overall, the presented results confirm that large-scale motions are self-sustained in the asymptotic suction boundary layer flow. LSMs in the boundary layer are not fed by smaller-scale active coherent structures near the wall. As all smaller scales are quenched, the energetic driving of LSMs only involves large scales on the order of the boundary layer thickness. The large-scale driving involves interactions with the mean velocity profile, indicating the importance of a correct mean profile for isolating LSMs without deteriorating them. 

\subsection{Dynamics of LSM in the large-scale minimal flow unit}\label{subsec:dynamics}
To investigate the dynamics of single, periodically replicated, LSMs and to study their self-sustained mechanism, we consider a computational domain that can accommodate a single LSM only, with the usual periodic boundary conditions in the horizontal plane. The overfiltered LES is thus carried out in the appropriate minimal flow unit for large-scale motions \citep{Hwang2010,Rawat2015a}. This horizontally-periodic domain, here referred to as the \textit{LSMbox}, is the large-scale equivalent of the minimal flow unit for near-wall turbulence \citep{Jimenez1991}.
In ASBL at $Re=300$, the size of the LSMbox is $\left(L_x=2.85\,\delta_{99}=192, L_z=1.25\,\delta_{99}=84\right)$. At the spatial resolution identical to previous overfiltered LES, $N_x=48,\ N_y=61,\ N_z=42$ collocation points are required for discretisation (see LSMbox in table \ref{tab:domains}).

\begin{figure}
    \centering
    \includegraphics[width=0.7\textwidth]{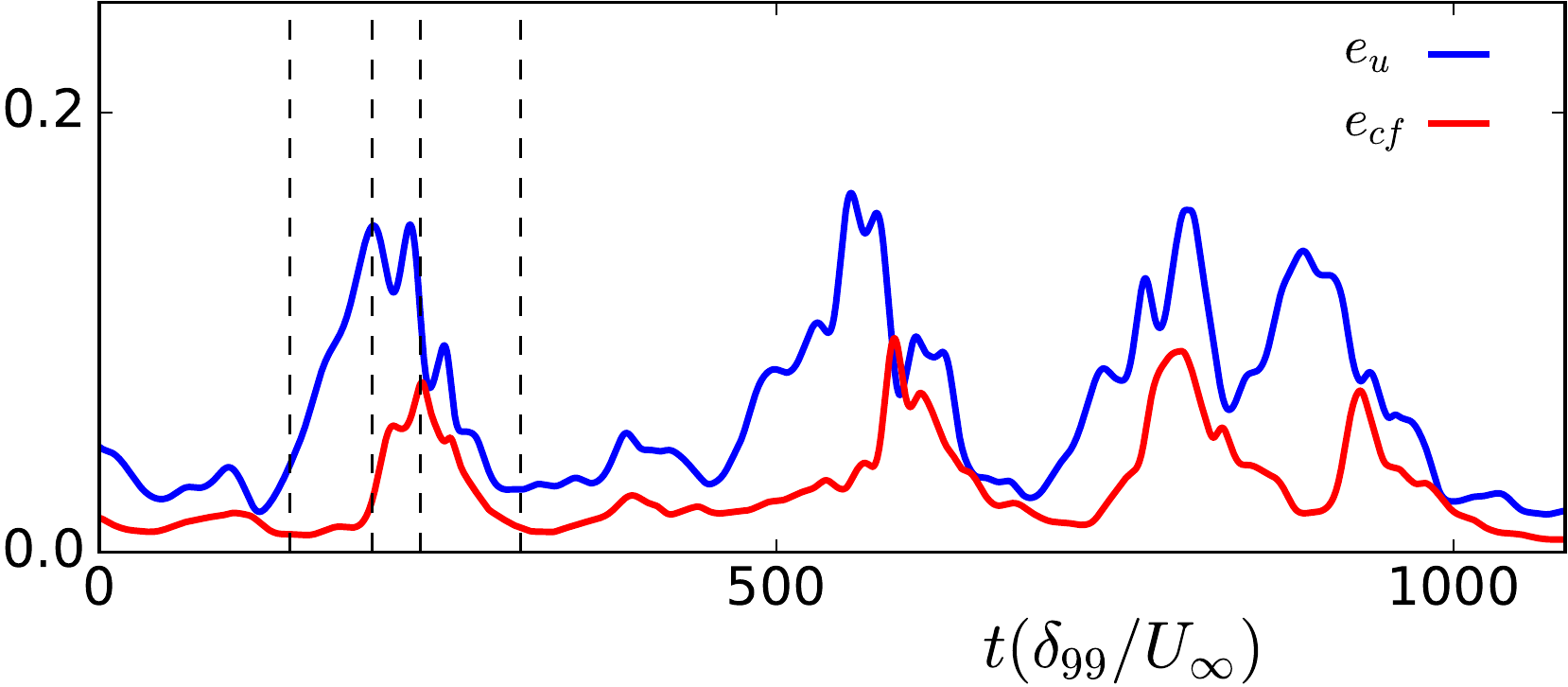}
    \caption{Temporal evolution of the streamwise kinetic energy per area of the wall $e_u=1/(L_x L_z)\int_{LSMbox} u^2 dx dy dz$ and the cross-flow kinetic energy per area of the wall $e_{cf}=1/(L_x L_z)\int_{LSMbox} \left(v^2 + w^2\right) dx dy dz$ for the overfiltered simulation with $C_s=0.3$ in the LSMbox. The vertical dashed lines indicate the times at which the snapshots of the flow are visualised in figure \ref{fig:LSM_vis}. 
    }
\label{fig:Eu_Ecf}
\end{figure}

\begin{figure}
    \centering
    ($a$)
    \includegraphics[width=0.2\textwidth]{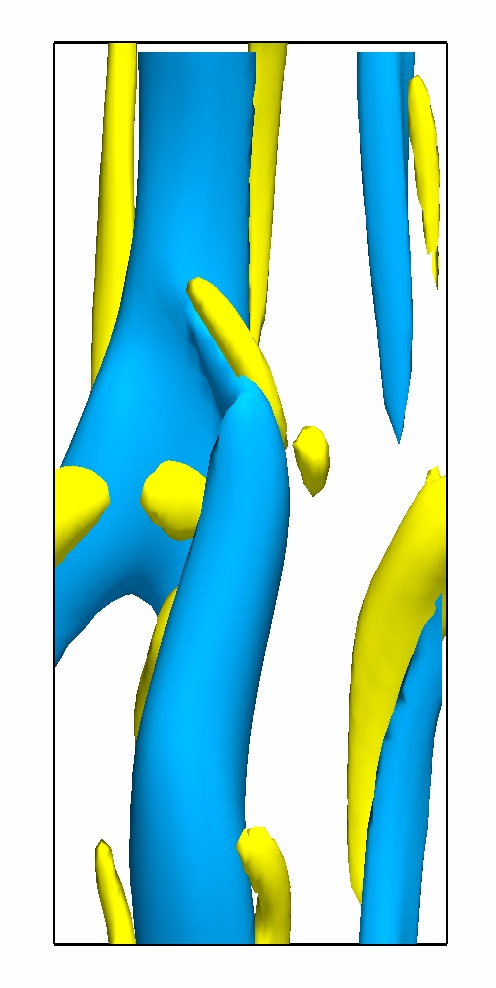}
    ($b$)
    \includegraphics[width=0.2\textwidth]{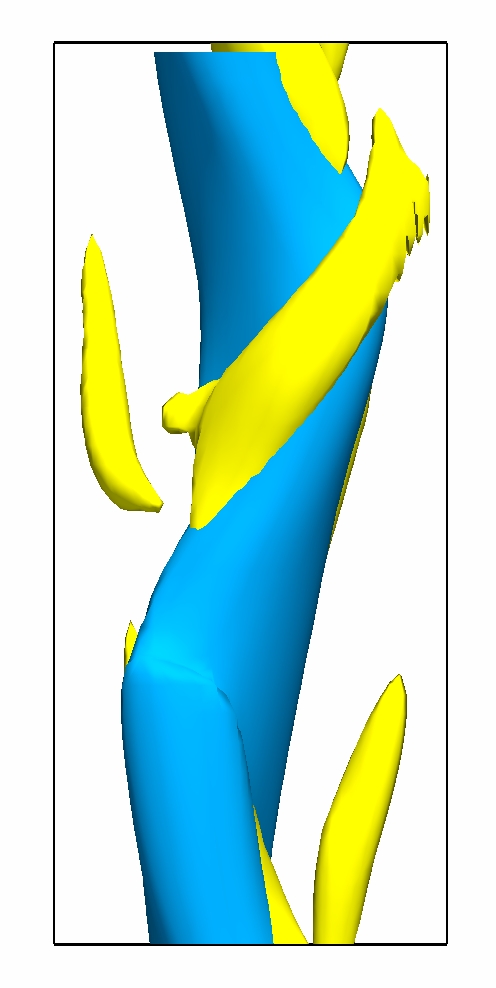}
    ($c$)
    \includegraphics[width=0.2\textwidth]{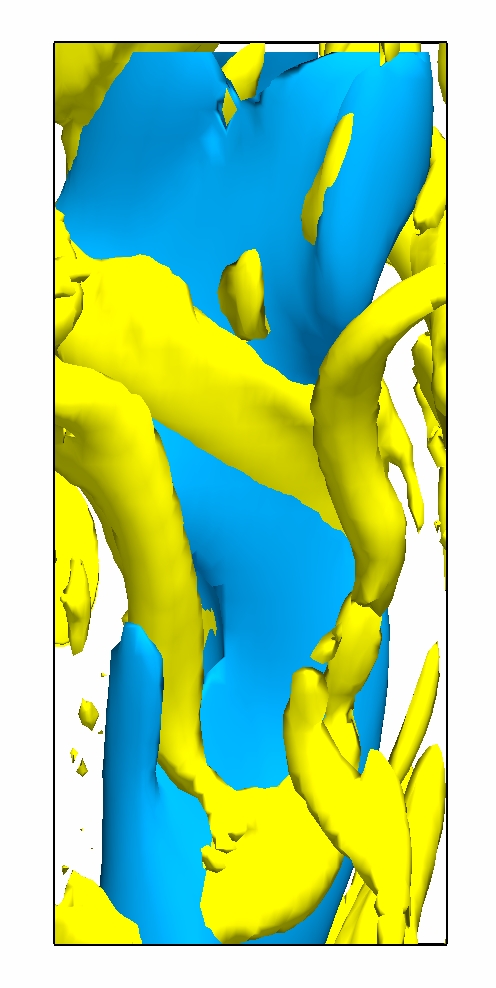}
    ($d$)
    \includegraphics[width=0.2\textwidth]{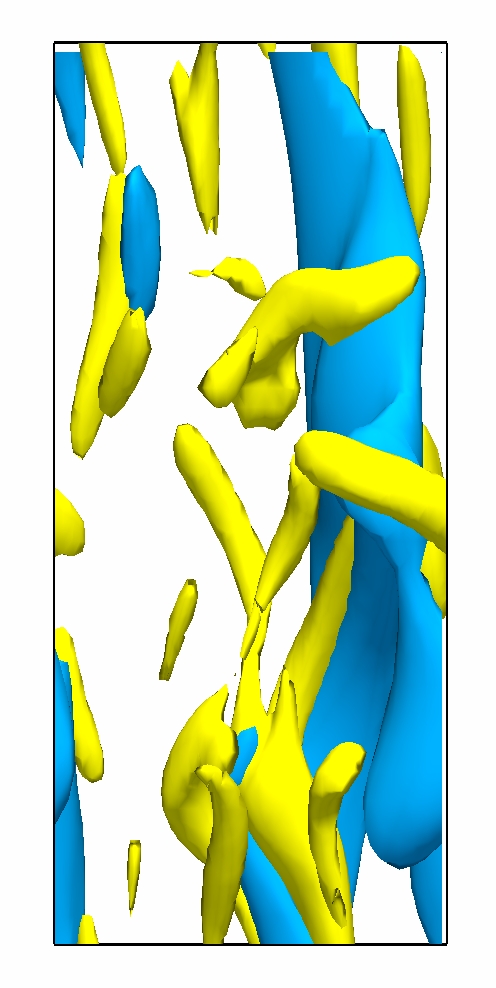}
    \caption{Snapshots of the overfiltered velocity field at times indicated in figure \ref{fig:Eu_Ecf}.
    The streak is visualised by the isosurface of the streamwise velocity $u^+=-0.52$ (blue), and vortices are visualised by the isosurface of the Q-criterion \citep{Jeong1995} with the iso-value of $3\%$ of the maximum (yellow). 
    The flow direction is upward and the view is oriented towards the wall.
    A large-scale motion is self-sustained by following the streak-vortices regeneration cycle: First, the downstream modulated streak, flanked by quasi-streamwise counter-rotating vortices, grows in amplitude (panels $a$, $b$ and $c$). The streak then becomes unstable and breaks down (panels $c$ and $d$). The resulting disorganises, vortices reorganise and the process repeats. 
    }
\label{fig:LSM_vis}
\end{figure}

Overfiltered LES in the minimal LSMbox for $C_s=0.3$ confirm that the single isolated LSM remains self-sustained also in the large-scale minimal flow unit (details are discussed in appendix \ref{sec:ODL_EMM_LSMbox}). This observation implies that the self-sustaining mechanism of LSMs is independent of even larger structures, that could in principle have been present in the larger LESbox studied above. Consequently, observations in ASBL agree with analogous observation in confined flows showing that LSMs are sustained independent of the dynamics of the small-scale structures in the near-wall region and independent of the presence of very-large-scale motions (VLSMs) \citep{Rawat2015a}.

In order to capture the dynamics of the self-sustained process we compute time series of the streamwise kinetic energy  
\begin{align*}
    e_u = 1/(L_x L_z)\int_{LSMbox} u^2 dx dy dz,
\end{align*}
which serves as a proxy for the intensity of streaks, and of the cross-flow kinetic energy
\begin{align*}
    e_{cf} = 1/(L_x L_z)\int_{LSMbox} \left(v^2 + w^2\right) dx dy dz,
\end{align*}
measuring the strengths of vortices.
Both the streamwise kinetic and the cross-flow kinetic energy exhibit an aperiodic bursting behaviour, as shown in figure \ref{fig:Eu_Ecf}.
While both components of energy evolve almost in phase, there is a small temporal phase-shift so that bursts of the streamwise kinetic energy $e_u$ slightly precede bursts of the cross-flow energy $e_{cf}$.
The evolution of quasi-streamwise streaks and vortices during one bursting cycle is visualised in figure \ref{fig:LSM_vis} where snapshots of the flow are shown. The visualisations of the flow fields demonstrate that the self-sustained process of LSMs involves the interaction of streaks and vortical structures. During a burst, sinuously bent streaks, that are flanked by quasi-streamwise counter-rotating vortices, grow in amplitude (panels $a$, $b$ and $c$), until they undergo breakdown (panels $c$ and $d$). Following breakdown the streaks and vortices are disorganised but they reorganise and the bursting process can start again.
The large-scale self-sustained bursting process thus strongly resembles processes observed in buffer-layer minimal flow units \citep{Jimenez1991}.

\section{Summary and conclusion}\label{sec:conclusion}

The goal of this study is to investigate if coherent large-scale motions (LSM) are self-sustained in boundary layers. Together with previous investigations of confined channel and Couette flows, the study of an open boundary layer aims at clarifying whether the LSM coherent self-sustained process is universally active in high-Reynolds-number turbulent wall-bounded flows.
To this end, the asymptotic suction boundary layer (ASBL) is chosen as testbed. The parallel nature of ASBL allows us to perform numerical simulations in a periodically continued domain and employ a filtering approach with properties that do not vary in stream- and spanwise direction. 
Since turbulent ASBL is dominated by the near-wall small-scale structures while large-scale motions are weaker than in other flow systems, ASBL is a particularly suitable system to study the self-sustained nature of large-scale motions.  Isolation of the weak large-scale motions in ASBL is a strong evidence that the self-sustained nature of LSMs is a universal property of turbulent wall-bounded flows. 

To determine whether LSMs are self-sustained or are driven by active motions at smaller scales we have built on the overfiltered large-eddy simulation (LES) approach where active small-scale structures are removed from the flow by increasing the LES spatial filter width while keeping a constant grid. This is realised by using the static \cite{Smagorinsky1963} model in `overfiltered' LES where the Smagorinsky constant $C_s$ is increased above its reference value best reproducing DNS statistics. 

Overfiltering attempts based on the original approach used in channel and Couette flows  \citep{Hwang2010,Hwang2011,Rawat2015a,Hwang2016a} in ASBL are inconclusive because the quantitative statistical properties of the isolated LSMs are affected at the large values of $C_s$ required to completely quench smaller-scale motions. 
To isolate LSMs we thus propose a modified overfiltering approach that preserves the original turbulent mean flow of the reference case. 
By using the modified method, the \emph{enforced mean model} EMM, we isolate the LSMs in ASBL. Thereby for the first time in an open boundary layer flow, we show that large-scale motions are indeed self-sustained even in the presence of active smaller-scale structures in the near-wall region.

Additional simulations in the large-scale minimal flow unit, containing a single LSM periodically replicated in the horizontal directions, show that the LSM self-sustaining mechanism does also not depend on the dynamics of larger scales.
The self-sustaining process of LSMs involves the aperiodic growth, breakdown and regeneration of sinuous streaks flanked by quasi-streamwise counter-rotating vortices.
This provides further evidence to the claim that wall-bounded turbulence can be associated with a continuum of self-sustained processes involving the mutual forcing and regeneration of coherent quasi-streamwise vortices and streaks with spatial scales ranging from those of buffer-layer structures to those of large-scale motions \citep{Hwang2011,Cossu2017}.

While a self-sustained process is generically observed in wall-bounded flows, its dynamical details appear to be non-universal at large scale.
In the ASBL, the observed large-scale process, that is associated with bursts in energy, is very similar to the one observed in buffer-layer minimal flow units \citep{Jimenez1991} but differs from the one observed for LSMs in channel and Couette flow \citep{Hwang2010,Rawat2015a}.
In channel and Couette flows the energies of large-scale streaks and quasi-streamwise vortices vary, most of the time, in phase opposition as in lower Reynolds number cases \citep{Hamilton1995, Waleffe1995}. In ASBL and the buffer-layer minimal flow unit, the energy of streaks and vortices however varies approximately in phase during aperiodic bursting events. 
This observation suggest that these two dynamically different behaviours are associated with situations where a single wall is dynamically relevant (boundary layers, buffer layers) versus those where two walls are relevant to the process (such as the plane channel or Couette flow). 
Current investigations are under way to compute invariant solutions and determine the structure of the phase space of coherent large-scale motions in the asymptotic suction boundary layer in order to elucidate the nature of the observed aperiodic bursting motions.

\section*{Acknowledgements}
This work was supported by the Swiss National Science Foundation (SNSF) under grant no. 200021-160088. SA acknowledges support by the State Secretariat for Education, Research and Innovation SERI via the Swiss Government Excellence Scholarship.

\section*{Declaration of interests}
The authors report no conflict of interest.


\appendix
\section{Self-sustained nature of LSMs in the absence of very-large-scale motions}\label{sec:ODL_EMM_LSMbox}

\begin{figure}
    \centering
    $(a)$
    \includegraphics[width=0.33\textwidth]{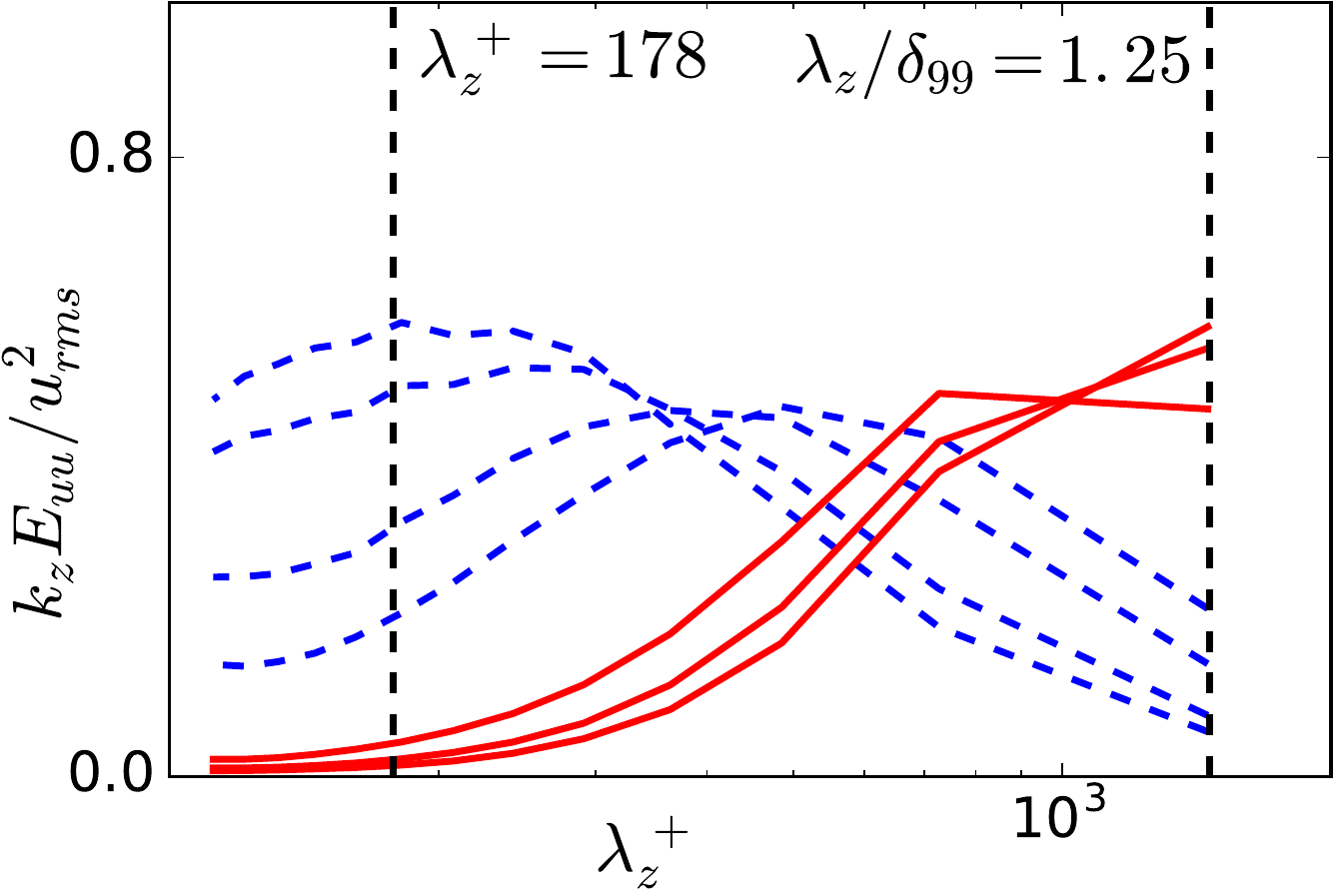}
    $(b)$
    \includegraphics[width=0.33\textwidth]{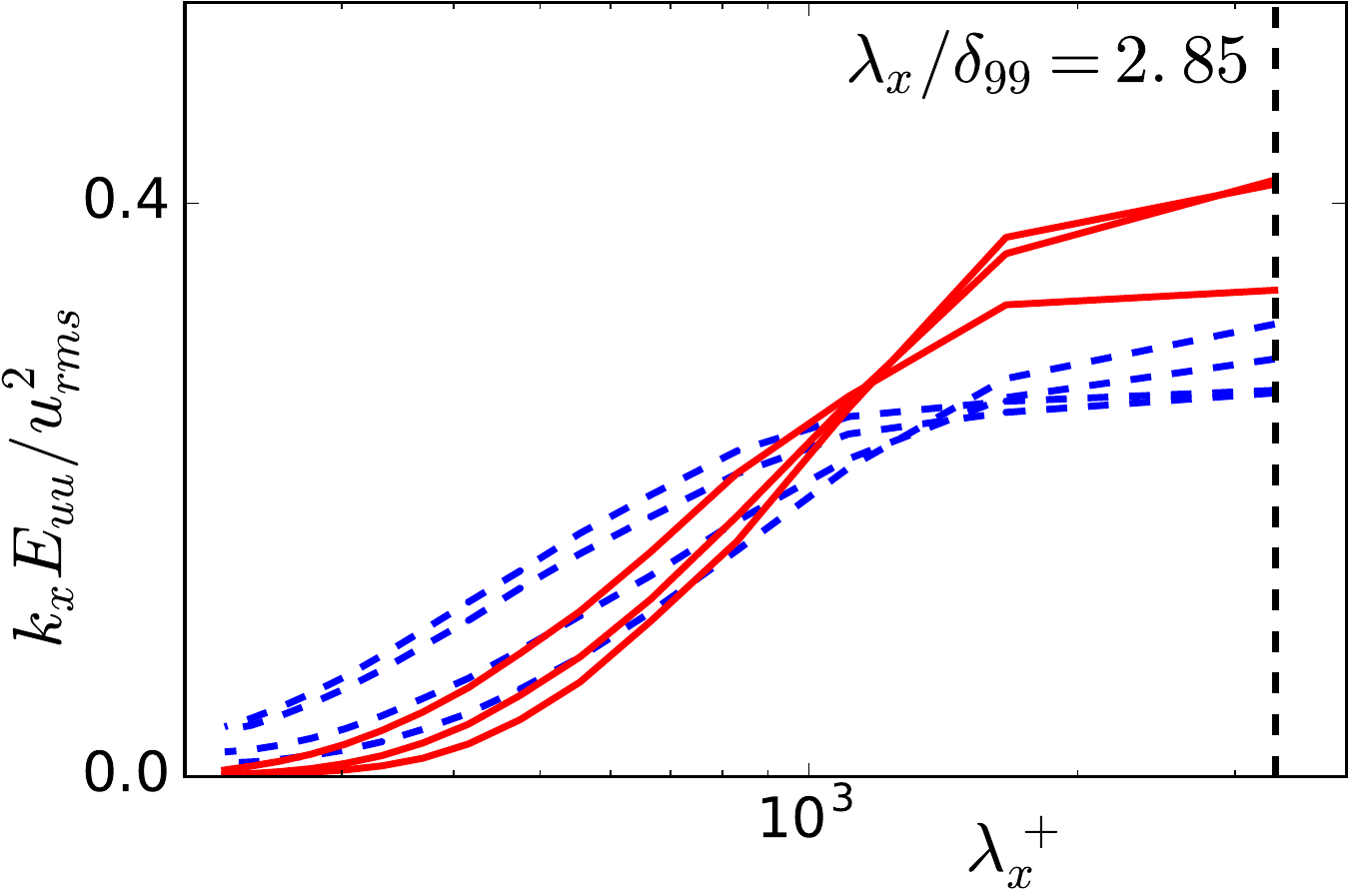}\\
    $(c)$
    \includegraphics[width=0.33\textwidth]{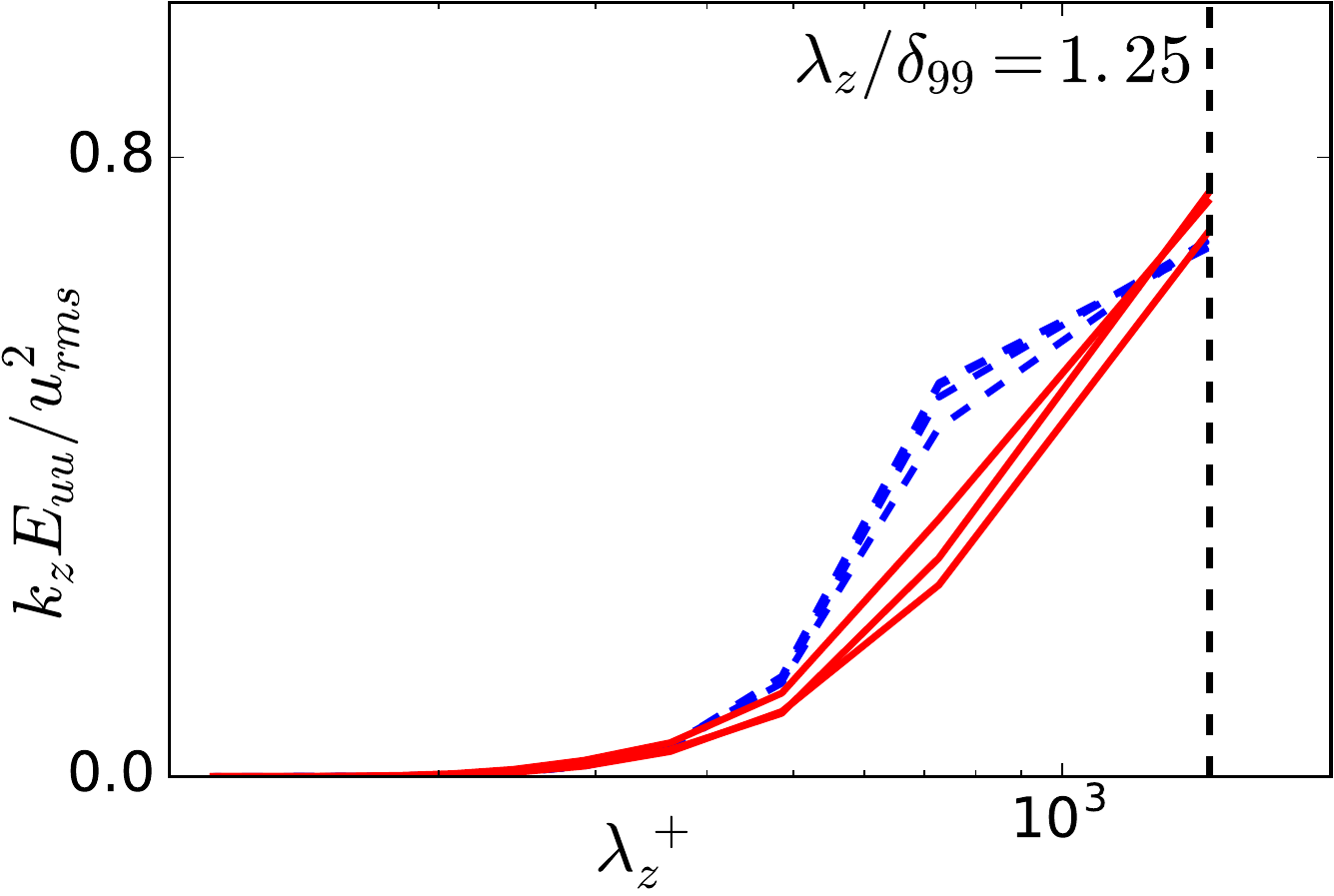}
    $(d)$
    \includegraphics[width=0.33\textwidth]{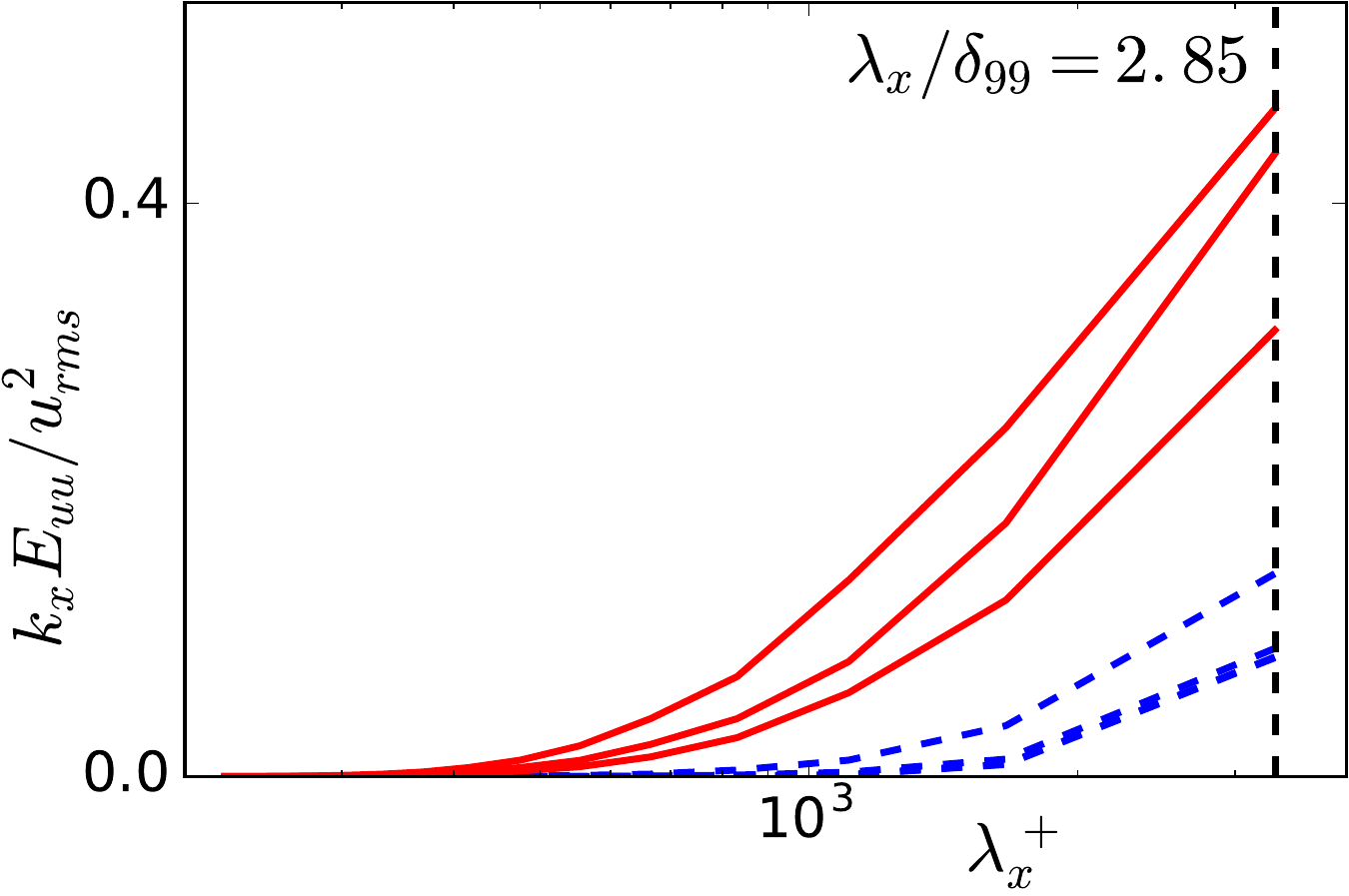}\\
    \caption{Spanwise premultiplied power spectra of the streamwise velocity (panels $a$ and $c$), and streamwise premultiplied power spectra of the streamwise velocity (panels $b$ and $d$) for simulations with the enforced mean model, EMM, in the LSMbox, at different values of $C_s$: $C_s = 0.045$ (panels $a$ and $b$); and $C_s = 0.3$ (panels $c$ and $d$). The data is shown for the inner region at $y^+ = \left[19, 30, 58, 94\right]$ (blue dashed lines) and for the outer region at $y/\delta_{99} = \left[0.3, 0.51, 0.74\right]$ (red solid lines).
    In the absence of very-large-scale motions, a single LSM survive when the smaller-scale structures are damped.}
\label{fig:ODL_EMM_LSMBox}
\end{figure}

Overfiltered simulations with enforced mean velocity profile are carried out in the LSMbox to determine if it is possible to isolate a single LSM from the dynamics of the small-scale structures in the absence of motions at scales larger than LSMs, such as the very-large-scale motions. 
Premultiplied power spectra at wall-normal sections in the near-wall region and in the outer region for simulations with the reference value $C_s=0.045$ and $C_s=0.3$ are shown in figure \ref{fig:ODL_EMM_LSMBox}.
At the reference value of Smagorinsky constant $C_s=0.045$, the spanwise and streamwise energy peaks corresponding to the near-wall small-scale structures (the peaks of the dashed blue lines) are present.
The energy peaks due to the single LSM are located at $\lambda_z=L_z$ and $\lambda_x=L_x$.
At $C_s=0.3$ (bottom panels), the near-wall energy peaks are damped while the LSM survives. 
Consequently, LSMs appear to be self-sustained when the small-scale structures are quenched by the filtering action and very-large-scale motions are eliminated by the periodic boundary conditions of the LSMbox.

\bibliographystyle{jfm_abbrv}
\bibliography{references}

\end{document}